\documentclass[fleqn,usenatbib]{mnras}
\usepackage{newtxtext,newtxmath}

\DeclareRobustCommand{\VAN}[3]{#2}
\let\VANthebibliography\thebibliography
\def\thebibliography{\DeclareRobustCommand{\VAN}[3]{##3}\VANthebibliography}

\usepackage{graphicx}	
\usepackage{amsmath}	
\usepackage{multicol}        
\usepackage{bm}		
\usepackage{pdflscape}	
\usepackage{fix-cm}
\usepackage[dvipsnames]{xcolor}

\defcitealias{Chatterjee2024}{Paper~I}

\newcommand{\dnu}{\Delta\nu}
\newcommand{\kpar}{k_{\parallel}}
\newcommand{\kpp}{k_{\perp}}
\newcommand{\kk}{{\bf k}}
\newcommand{\V}{\mathcal{V}}
\newcommand{\vcg}{\mathcal{V}_{cg}}
\newcommand{\U}{\mathbf{U}}
\newcommand{\HI}{\ion{H}{i}}
\newcommand{\eg}{{e.g.~}}
\newcommand{\cl}{C_{\ell}}
\newcommand{\nub}{\bar{\nu}}

\newcommand{\pk}{P(k_{\perp}, k_{\parallel})}
\newcommand{\pkm}{P^m(k_{\perp}, k_{\parallel})}
\newcommand{\n}{\hat{\mathbf{n}}}

\usepackage[T1]{fontenc}
\usepackage{ae,aecompl}

\usepackage{newtxtext,newtxmath}

\title[TTGE II: The Missing Frequency Channels]{The Tracking Tapered Gridded Estimator for the 21-cm power spectrum from MWA drift scan observations II: The Missing Frequency Channels}

\author[A. Elahi et al.]{Khandakar Md Asif Elahi,$^{1, 2}$\thanks{E-mail:asifelahi999@gmail.com}
Somnath Bharadwaj,$^{2}$\thanks{E-mail:somnath@phy.iitkgp.ac.in}
Suman Chatterjee,$^{3}$
Shouvik Sarkar,$^{1}$
\newauthor
Samir Choudhuri,$^{1}$
Shiv Sethi,$^{4}$
Akash Kumar Patwa,$^{4}$
\\
\\
$^{1}$ Centre for Strings, Gravitation and Cosmology, Department of Physics, Indian Institute of Technology Madras, Chennai 600036, India\\
$^{2}$ Department of Physics, Indian Institute of Technology Kharagpur, Kharagpur - 721302, India\\
$^{3}$ Department of Physics and Astronomy, University of the Western Cape, 7535 Bellvill, Cape Town, South Africa\\
$^{4}$ Raman Research Institute, C. V. Raman Avenue, Sadashivanagar, Bengaluru 560080, India\\
}

\date{Accepted XXX. Received YYY; in original form ZZZ}

\pubyear{\the\year{}}

\begin{document}
\label{firstpage}
\pagerange{\pageref{firstpage}--\pageref{lastpage}}
\maketitle

\begin{abstract} 
Missing frequency channels pose a problem in estimating the redshifted 21-cm power spectrum $P(k_\perp,k_\parallel)$ from radio-interferometric visibility data. This is particularly severe for the Murchison Widefield Array (MWA), which has a periodic pattern of missing channels that introduces spikes along $k_\parallel$. The Tracking Tapered Gridded Estimator (TTGE) overcomes this by first correlating the visibilities in the frequency domain to estimate the multi-frequency angular power spectrum (MAPS) $C_\ell(\Delta\nu)$ that has no missing frequency separation $\Delta\nu$. We perform a Fourier transform along $\Delta\nu$ to estimate $P(k_\perp,k_\parallel)$. Simulations demonstrate that the TTGE can estimate $P(k_\perp,k_\parallel)$ without any artifacts due to missing channels. However, the spikes persist for the actual foreground-dominated data. A detailed investigation, considering both simulations and actual data, reveals that the spikes originate from a combination of the missing channels and the strong spectral dependence of the foregrounds. We propose and demonstrate a technique to mitigate the spikes. Applying this, we find the values of $P(k_\perp,k_\parallel)$ in the region $0.004 \leq  k_\perp \leq 0.048\,{\rm Mpc^{-1}}$ and $k_\parallel > 0.35 \,{\rm Mpc^{-1}}$ to be consistent with zero within the expected statistical fluctuations. We obtain the $2\sigma$ upper limit of $\Delta_{\rm UL}^2(k)=(934.60)^2\,{\rm mK^2}$  at $k=0.418\,{\rm Mpc^{-1}}$ for the mean squared  brightness temperature fluctuations of the $z=8.2$ epoch of reionization (EoR) 21-cm signal. This upper limit is from  $\sim17$ minutes of observation for a single pointing direction. We expect tighter constraints when we combine all $162$  different pointing directions of the drift scan observation.
\end{abstract}




\section{Introduction}
\label{sec:intro}  
The study of the epoch of reionization (EoR), when the neutral hydrogen (\HI) in the diffuse inter-galactic medium (IGM) first became ionized, is of considerable scientific interest.  There are  several observational efforts currently underway to measure the power spectrum (PS) of the brightness temperature fluctuations of the  redshifted  \HI~ 21-cm signal, which is one of the most promising direct probes of EoR. Several radio interferometers including  the  Murchison Widefield Array (MWA; 
\citealt{Tingay2013}), LOw Frequency ARray (LOFAR; \citealt{vanHarlem2013}), 
Hydrogen Epoch of Reionization Array (HERA; \citealt{Deboer2017}),  Giant Metrewave Radio Telescope (GMRT; \citealt{Swarup1991, Gupta2017}) and the upcoming SKA-low \citep{Mellema2013, Koopmans2015} all target to detect this signal.  This EoR 21-cm PS, however, still remains to be detected  and we currently only have upper limits  \citep{Paciga2013, Kolopanis2019, Mertens2020, Trott2020, Pal2020, Abdurashidova2022, Kolopanis2023, Acharya2024}. The best upper limit at present is   $\Delta^2(k) < (30.76)^{2} \, {\rm mK}^2$  at $k = 0.192\, h\, {\rm Mpc}^{-1}$  for  $z = 7.9$ from HERA \citep{Abdurashidova2022}.

Foregrounds that are three to four orders of magnitude brighter 
pose the biggest challenge for detecting the EoR 21-cm PS 
\citep{Ali2008, Bernardi2009, Ghosh2012, Paciga2013, Patil2017}. 
Several approaches have been developed to overcome this issue.  Most of these 
rely on the fact that foregrounds are spectrally smooth compared to the 21-cm signal. One approach, namely `foreground removal', attempts to subtract out the foregrounds (e.g. \citealt{Chapman2012, Mertens2018, Elahi2023b}). Another approach, namely  `foreground avoidance', only uses the $(k_\perp, \kpar)$ 
modes outside the `foreground wedge'  \citep{Datta2010, Morales2012, Vedantham2012, Trott2012, Pober2016} to estimate  21-cm PS (e.g. \citealt{Dillon2014, Dillon2015b, Trott2020, Pal2020, Pal2022, Abdurashidova2022, Elahi2023}). 
The success of these techniques, however, relies on several important steps, including the flagging of corrupted data \citep{Offringa2015, Wilensky2019_SSINS}, precise calibration of the instrument \citep{Barry16, Byrne2019, Kern2020, Gan2023, Gayen2025}, and accurate modelling and subtraction of both extragalactic point sources \citep{Ali2008, Ceccotti2025} and the diffuse Galactic foregrounds \citep{Choudhuri2017, Byrne2022, Gehlot2022dgse}.

The Tapered Gridded Estimator (TGE; \citealt{Choudhuri2014, Choudhuri2016b}) is a visibility-based 21-cm PS estimator that suppresses the sidelobe responses of the telescope to mitigate the effects of extragalactic point source foregrounds \citep{Ghosh2011a, Ghosh2011b}. The TGE has been extensively used for measuring the 21-cm PS \citep{Pal2020, Pal2022, Elahi2023, Elahi2023b, Elahi2024}.
In a recent work \cite{Chatterjee2022} have introduced the Tracking Tapered Gridded Estimator (TTGE), which generalises the TGE for estimating the 21-cm PS from drift scan observations. The present paper is the second in a series of papers that apply the   TTGE  to constrain the EoR 21-cm PS using MWA drift scan observations.

Missing frequency channels pose a serious problem for visibility-based  PS estimation. In `delay space analysis', the measured visibilities are Fourier transformed along frequency to estimate the delay space visibilities, which are then used to estimate the PS \citep{Morales2004, Parsons2012b}. In this approach, the missing frequency channels introduce ringing artifacts in the delay space visibilities and corrupt the estimated PS. \cite{Parsons2009} introduced a one-dimensional (1D) CLEAN algorithm to get uncorrupted delay space visibilities from RFI-contaminated data. The 1D CLEAN is an adaptation of the two-dimensional \textsc{CLEAN} algorithm used in aperture synthesis \citep{Hogbom1974, Roberts1987}. It performs a nonlinear deconvolution in the delay space, which is equivalent to a least-squares interpolation in the frequency domain \citep{Roberts1987}. It has been extensively used in 21-cm PS experiments \citep{Parsons2014, Chakraborty2021, Abdurashidova2022}. 
\cite{Trott2016a} used the Least Square Spectral Analysis (LSSA) technique to deal with the irregular and missing data in the visibilities while transforming from frequency to delay space. LSSA has been used for estimating the EoR 21-cm PS from MWA \citep{Trott2020} and LOFAR \citep{Patil2017, Mertens2020}. \cite{Chakraborty2022} have used both simulated and actual data to make a detailed comparison of these two methods. $\mathcal{E}$ppsilon \citep{Barry2019eppsilon} and DAYENU \citep{Ewall-Wice2021} are two comparatively newer methods which try to fill in or `inpaint' the missing data in the visibilities before doing a Fourier transform. Recently, \cite{Chen2025} have investigated the effects of inpainting for HERA. Several other algorithms, such as Gaussian Process Regression (GPR; \citealt{Mertens2018, Mertens2020, Trott2020, Kern2021}), Gaussian Constrained Realizations (GCR; \citealt{Kennedy2023}) etc., have also been used for accurate estimation of the $21$-cm PS from data that have missing channels.

The missing channels do not pose a problem if we first correlate the visibilities in the frequency domain \citep{Bharadwaj2001a, Bharadwaj2005} to estimate the multi-frequency angular power spectrum (MAPS) $C_\ell(\Delta\nu)$, and then perform a Fourier transform along $\Delta\nu$ to estimate the 21-cm PS $P(k_\perp,k_\parallel)$. Even if there are many frequency channels $\nu$ missing, the estimated $C_\ell(\Delta\nu)$ does not have a missing frequency separation $\Delta\nu$. Therefore, in this approach, the power spectrum does not show artifacts due to the missing channels. The key idea is that it is not essential to compensate for the missing frequency channels, as the power spectrum can be estimated using only the available channels. The TGE and the TTGE both incorporate this idea.   \cite{Bharadwaj2018} have used simulations to show it is possible to apply the TGE to estimate $P(k_\perp, \kpar)$ without any artifacts even when $80 \%$ of randomly chosen frequency channels are flagged. This TGE is also found to perform well on actual data \citep{Pal2020, Pal2022, Elahi2023, Elahi2023b, Elahi2024}.

The MWA has a periodic pattern of flagged channels, for which the delay space analysis 
introduces horizontal streaks in  $P(k_\perp, \kpar)$  \citep{Paul2016, Li2019, Trott2020, Patwa2021}.  In \cite{Chatterjee2024} (hereafter \citetalias{Chatterjee2024}), which is the first paper in this series, we have first validated the TTGE using simulations of the drift-scan MWA observations. These incorporate exactly the same flagging as the actual MWA data. For a three dimensional (3D) cosmological signal, the TTGE is able to recover $\pk$ without any artifacts due to the flagged channels. In \citetalias{Chatterjee2024}, we also present preliminary results for the actual MWA data considering  a single pointing at $({\rm RA,DEC)} = (6.1^{\circ},-26.7^{\circ})$.   For the actual data that is foreground dominated, considering $\pk$,  we find a periodic pattern of spikes along $\kpar$. The period of the spikes corresponds to $1.28 \, {\rm MHz}$, which is the period of the pattern of flagged channels. This manifests itself as streaks in the $(\kpp,\kpar)$ plane. Although the amplitude of these artifacts is much smaller compared to those in the delay-space analysis, it suffices to contaminate a large region of the $(\kpp,\kpar)$ plane, and 21-cm PS estimation  is restricted to a small  rectangular region ($0.05 \leq \kpp \leq 0.16 \, {\rm Mpc^{-1}}$, $0.9 \leq \kpar \leq 4.6 \, {\rm Mpc^{-1}}$)  where $\mid P(\kpp, \kpar) \mid $ is found to have values in the range $10^7 - 10^{11} \, {\rm mK^2}\, {\rm Mpc^3}$. We use this to place a $2\sigma$ upper limit $\Delta_{UL}^2(k) = (1.85\times10^4)^2\, {\rm mK^2}$ on the mean squared 21-cm brightness temperature fluctuations at $k=1 \,{\rm Mpc}^{-1}$. 

In the present paper, we investigate the effect of missing channels on power spectrum estimation,  particularly focusing on the foregrounds that dominate the actual MWA data.  The entire analysis is restricted to the single pointing that was considered in \citetalias{Chatterjee2024}. Based on our analysis, which covers simulations in addition to the actual data, we propose a method to mitigate the artifacts that arise due to the missing channels. We have applied this method to obtain improved constraints on the 21-cm PS. A brief outline of the paper follows. Section~\ref{sec:data} presents a brief description of the data we have analyzed in this paper. In Section~\ref{sec:MAPS}, we briefly review our method of power spectrum estimation, and in Section~\ref{sec:mwaflag}, we present a detailed analysis of how the missing channels affect power spectrum estimation from the actual MWA data. In Section~\ref{sec:simflag}, we use simulations to study the impact of missing channels, and in Section~\ref{sec:SCF}, we propose Smooth Component Filtering (SCF) as a method to mitigate the artifacts produced by the missing channels. Finally, we present the results in Section~\ref{sec:MWA}, and the summary and conclusions in Section~\ref{sec:summary}.  In Appendix~\ref{sec:sim} we present realistic simulations that validate SCF for both the expected 21-cm signal and foregrounds. In Appendix~\ref{sec:window}, we present the rationale for the choice of a window function in SCF.

The actual implementation of the TTGE that is used to estimate the 21-cm PS from the measured visibilities closely follows \citetalias{Chatterjee2024}, and we have not explicitly discussed the details here. For the cosmological parameters, we have used the values from \cite{Planck2020f}.

\section{Data Description}
\label{sec:data}

The MWA (\citealt{Lonsdale2009}, \citealt{Wayth2018}) Phase~II drift scan observation considered here (project ID G0031) is described in detail in \citet{Patwa2021}, and we briefly describe it here for completeness. The observation is carried out at a fixed declination (DEC)  $-26.7^{\circ}$ which corresponds to the zenith. The drift scan covers the right ascension (RA) range $349^{\circ}$ to $70.3^{\circ}$ (total $81.3^{\circ}$) over a time duration of  5~hr 24~min. The visibility data are recorded every 2~min,
which results in $162$ different pointing centers (PCs), located at an interval of $0.5^{\circ}$ along RA. We label these PCs as PC=1,2,...,162. The observation was carried out on 10 consecutive nights.  The observation has been performed at the central frequency of $\nu_c=154.2 \, {\rm MHz}$ with $N_c = 768$ channels of resolution of $\dnu_c=40 \, {\rm kHz}$  covering the total observing bandwidth of $B_{\rm bw} = 30.72$ MHz. This is further divided into  24 coarse bands each containing  32 channels or $1.28 \, {\rm MHz}$.

The data has been pre-processed with COTTER (\citealt{Offringa2015}) which flags RFI and non-working antennas. Considering each coarse band, it also flags four channels at both ends and one channel at the center, which results in channels (1-4,17,29-32) to be flagged. This particular flagging is done to avoid leakages of signal from the adjacent coarse bands (\citealt{Prabu2015}). COTTER is applied individually to the 0.5~s time resolution visibility data, and then the data are averaged to 10~s time resolution. This pre-processed data are written in \textit{Measurement Sets} (MS), which are readable in \texttt{CASA}\footnote{\url{https://casa.nrao.edu/}} (\citealt{casa07}). The  MS for 10 nights are calibrated separately using the strong and unresolved calibrator source Pictor~A (\citealt{Patwa2021}). 

We note that the first $\sim2 \, {\rm hr}$ of data are missing from the $6$-th night, and consequently the total nights of observations $(N_{\rm nights})$ for a PC is either 9 or 10.  We perform Local Sidereal Time (LST) stacking (\eg \citealt{Bandura2014, Amiri2022}) over the data from different nights and obtain the equivalent of one night of drift scan data. Therefore, we finally have $162$ MS,  each of which corresponds to a different pointing direction on the sky, and contains visibility data with $11$ different time stamps each with $t_{\rm int}= N_{\rm nights} \times 10~{\rm s}$ effective integration time. 

In this paper, the analysis is restricted to a single pointing PC=34 $({\rm RA} = 6.1^{\circ})$ which has also been analyzed in \citetalias{Chatterjee2024}. For this PC, we have $N_{\rm nights}=9$, and the r.m.s. system noise for the real (and imaginary) part of the visibility data is 20~Jy (eq. 1 of \citetalias{Chatterjee2024}). Following \citetalias{Chatterjee2024}, we have restricted the entire analysis to the baseline range $6\lambda \leq \mid U \mid \leq 220 \lambda$.

\section{Power Spectrum Estimation from MAPS}
\label{sec:MAPS}
Our estimator  is based on the formula \citep{Bharadwaj2001b,Bharadwaj2005}
\begin{equation}
    \langle  \V(\U,\nu) \V^{*}(\U,\nu+ \Delta \nu) \rangle = \left[\frac{Q^2 \theta_0^2}{4 r^2} \right]  \int_{-\infty}^{\infty} d k_{\parallel} \, \, e^{i \, k_{\parallel} r^{'}  \Delta \nu} 
\, P(\kk),
\label{eq:pk1}
\end{equation}
which relates the correlation between two visibilities measured at the same baseline $\U$  but slightly different frequencies to the redshifted 21-cm power spectrum  (PS) $P(\kk)$, where $\kk$ has components ${\mathbf k_{\perp}}=\frac{2 \pi \U}{r}$ and  $k_{\parallel}$ respectively perpendicular and parallel to the line of sight (LoS). Here, $r$ is the comoving distance to the \HI{} from which the redshifted 21-cm radiation  is received at $\nu$, $ r^{'}=\frac{d r}{d \nu}$, $Q=2 k_B/\lambda^2$  and 
\begin{equation}
    \left(\frac{\theta_0}{2}\right)^2=\frac{1}{2 \pi} \int d^2 \theta \, \mid \mathcal{A}(\theta) \mid^2 \,. 
    \label{eq:bm1} 
\end{equation}
where $\mathcal{A}(\theta)$ is the telescope's primary beam (PB) response at an angle $\theta$ away from the phase center of the observation. 
If we assume that the PB can be well approximated by a Gaussian, we have $\mathcal{A}(\theta) \approx e^{-\theta^2/\theta_0^2}$ and $\theta_0 \approx 0.6 \, \theta_{\rm FWHM}$.

We have implemented this in two steps. In the first step, we use  
\begin{equation}
    C_{\ell}( \nu_a,\nu_b)  = \left[\frac{2}{\pi Q^2 \theta_0^2} \right]_{\nu_c}  \langle  \V(\U,\nu_a) \V(\U,\nu_b) \rangle \, 
    \label{eq:maps1}
\end{equation}
to estimate $C_{\ell}( \nu_a,\nu_b)$, the multi-frequency angular power spectrum (MAPS), with  $ \ell = 2 \pi \mid \U \mid$. We assume that the bandwidth of observations is small, and the quantities within the square brackets $[...]$  can all be evaluated at the central frequency $\nu_c$. Further, the 21-cm signal is assumed to be ergodic along the LoS (see \citealt{Mondal2018} for details),  whereby the MAPS depends only on the frequency separation, and not the individual frequencies 
\begin{equation}
    C_{\ell}(\Delta \nu) \equiv C_{\ell}(\mid \nu_a-\nu_b \mid)=C_{\ell}(\nu_a,\nu_b) \,.
    \label{eq:maps2|}
\end{equation}

In the second step we use \citep{Datta2007, Mondal2018}
\begin{equation}
    \pk = r^2 r^{'} \int_{-\infty}^{\infty} d(\Delta \nu) \, e^{-i \, k_{\parallel} r^{'} \Delta \nu} C_{\ell}(\Delta \nu) 
    \label{eq:pk2} 
\end{equation}
to estimate the cylindrical PS $\pk$. 

The actual implementation of TTGE (\citealt{Chatterjee2022}, \citetalias{Chatterjee2024}) utilizes 
\begin{equation}
 C_{\ell_g}( \nu_a,\nu_b)=M^{-1}_g(\nu_a,\nu_b) \,  
       \langle  \vcg(\nu_a) \vcg^*(\nu_b) \rangle \,
    \label{eq:maps3}
\end{equation}
to estimate $C_{\ell_g}( \nu_a,\nu_b)$  instead of  eq.~(\ref{eq:maps1}). Here, $\V_{cg}(\nu)$ refers to the convolved, gridded visibilities that are evaluated on a rectangular grid in the `$u-v$' plane,  and $g$ refers to a particular grid point with corresponding baseline $\U_g$ and angular multipole $\ell_g$. $M_g(\nu_a,\nu_b)$ is a normalization factor whose value is estimated using simulations.  
Note that we consider the convolved, gridded visibilities $\vcg(\nu)$ for two different polarizations $(XX, YY)$, and cross-correlate these to 
estimate $C_{\ell_g}( \nu_a,\nu_b)$ in eq.~(\ref{eq:maps3}). This enables us to 
avoid the noise bias that appears if we correlate a visibility with itself. The details of the cross-correlation estimator are presented in \cite{Elahi2023}.

For the purpose of the discussion in  Section~\ref{sec:simflag}, where we consider a fixed grid point $g$,  and analyze the effect of flagged frequency channels on PS estimation using simulated data,  it suffices to set $M_g(\nu_a,\nu_b)$ to a constant value  $M_g(\nu_a,\nu_b)=1$.  Note that we have used the full implementation of TTGE, as described in  \citetalias{Chatterjee2024}, to analyze the actual MWA data for which the results are presented in Sections~\ref{sec:mwaflag}, \ref{sec:SCF} and \ref{sec:MWA}. 

\section{Effect of the Flagged Channels on MWA data}
\label{sec:mwaflag}

The present observation  covers a frequency  bandwidth of $B_{\rm bw} = 30.72$ MHz, with $N_c = 768$ channels of resolution  $\dnu_c=40 \, {\rm kHz}$. The band is further divided into  24 coarse bands, each containing  32 channels that span  $1.28 \, {\rm MHz}$. The first four, last four, and the central channel of each coarse band are flagged. This introduces a periodic pattern of flagged channels in the MWA data. In addition, several other channels may be flagged to avoid RFI or for various other reasons. In this section, we analyze the impact of flagged channels on power spectrum estimation. 
Note that the entire analyses of Sections~\ref{sec:mwaflag}, ~\ref{sec:simflag} and ~\ref{sec:SCF} are restricted to a single grid point $g$, which corresponds to $\ell = 142$. However, we omit the subscript $g$ in $\ell_g$ in these sections to simplify the notation.


\begin{figure}
   \includegraphics[width=\columnwidth]{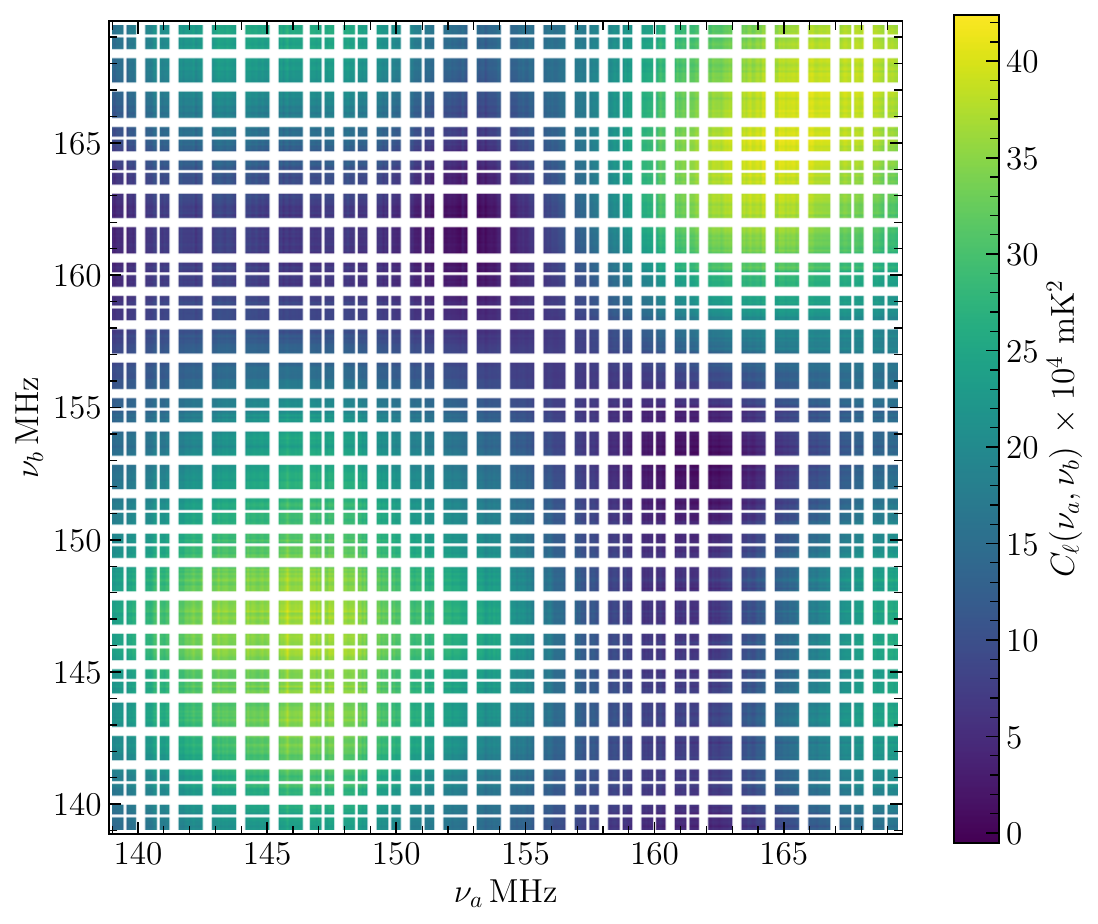}
    \caption{The MAPS $C_{\ell}(\nu_a,\nu_b)$ for the MWA data considering a particular grid point corresponding to $\ell = 142$.}
    \label{fig:mwa-maps1}
\end{figure}
  
Figure~\ref{fig:mwa-maps1} shows the MAPS $C_{\ell}(\nu_a,\nu_b)$ for the MWA data. Note the pattern of missing estimates due to the periodic pattern of flagged channels. Although there are many pairs of frequency channels $(\nu_a,\nu_b)$ that are missing, we do not have any missing $C_{\ell}(\Delta \nu)$ within the allowed range of 
$\Delta \nu$ values. To analyze this better, it is useful  to introduce the variables 
\begin{equation}
    \nub=(\nu_a+\nu_b)/2  \hspace{0.5in}  {\rm and} \hspace{0.5in}  \dnu= \nu_a-\nu_b \,,
\end{equation}
and analyze  $C_{\ell}(\dnu,\nub)$ instead of $C_{\ell}(\nu_a,\nu_b)$.  Considering $C_{\ell}(\dnu,\nub)$, we may interpret that the $\dnu$ dependence quantifies how the signal decorrelates as we change the frequency separation, whereas the $\nub$ dependence quantifies the overall spectral dependence of the signal.  Note that we do not expect any $\nub$ dependence ({\it ie.} $C_{\ell}(\dnu)=C_{\ell}(\dnu,\nub)$) for a statistically homogeneous,  3D  signal that is ergodic along the LoS.

\begin{figure}
   \includegraphics[width=\columnwidth]{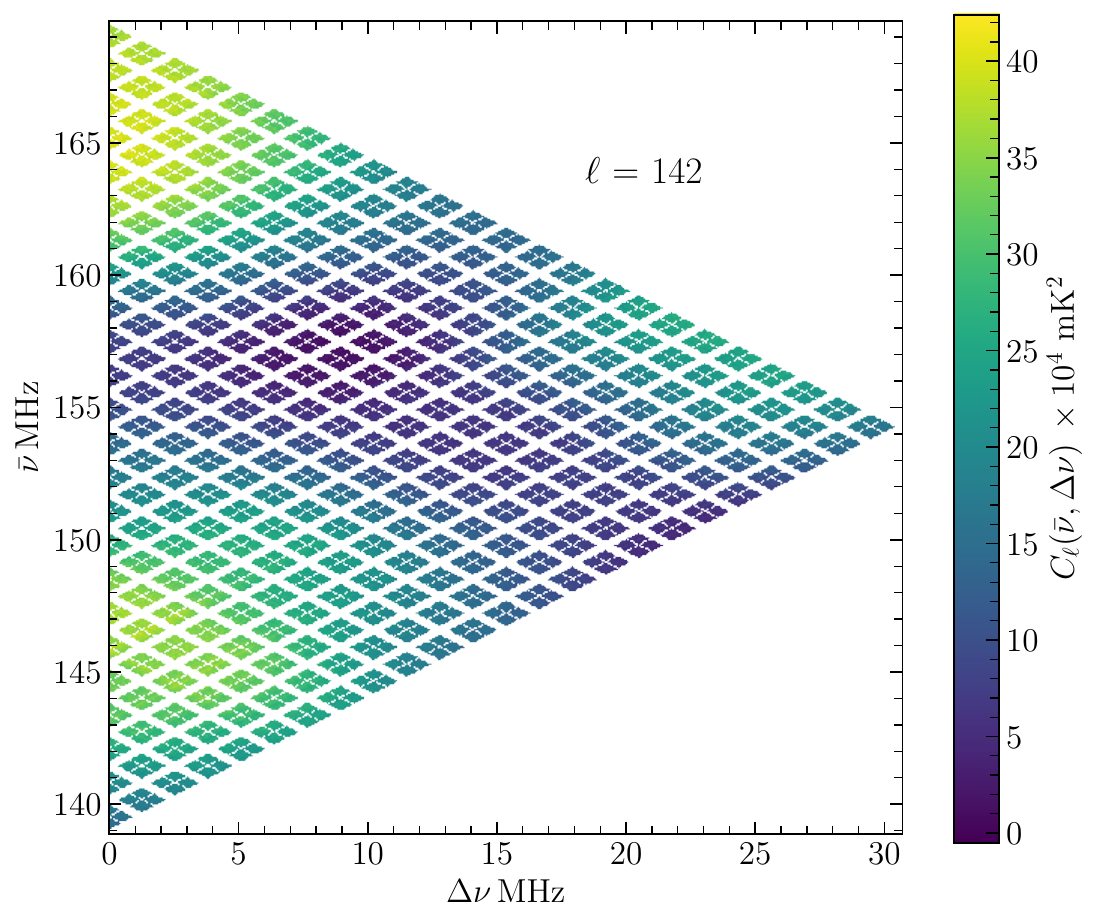}
    \caption{The MAPS $C_{\ell}(\dnu,\nub)$ corresponding to the MAPS $C_{\ell}(\nu_a,\nu_b)$ shown in Figure~\ref{fig:mwa-maps1}.}
    \label{fig:maps_nubar}
\end{figure}

\begin{figure}
   \includegraphics[width=\columnwidth]{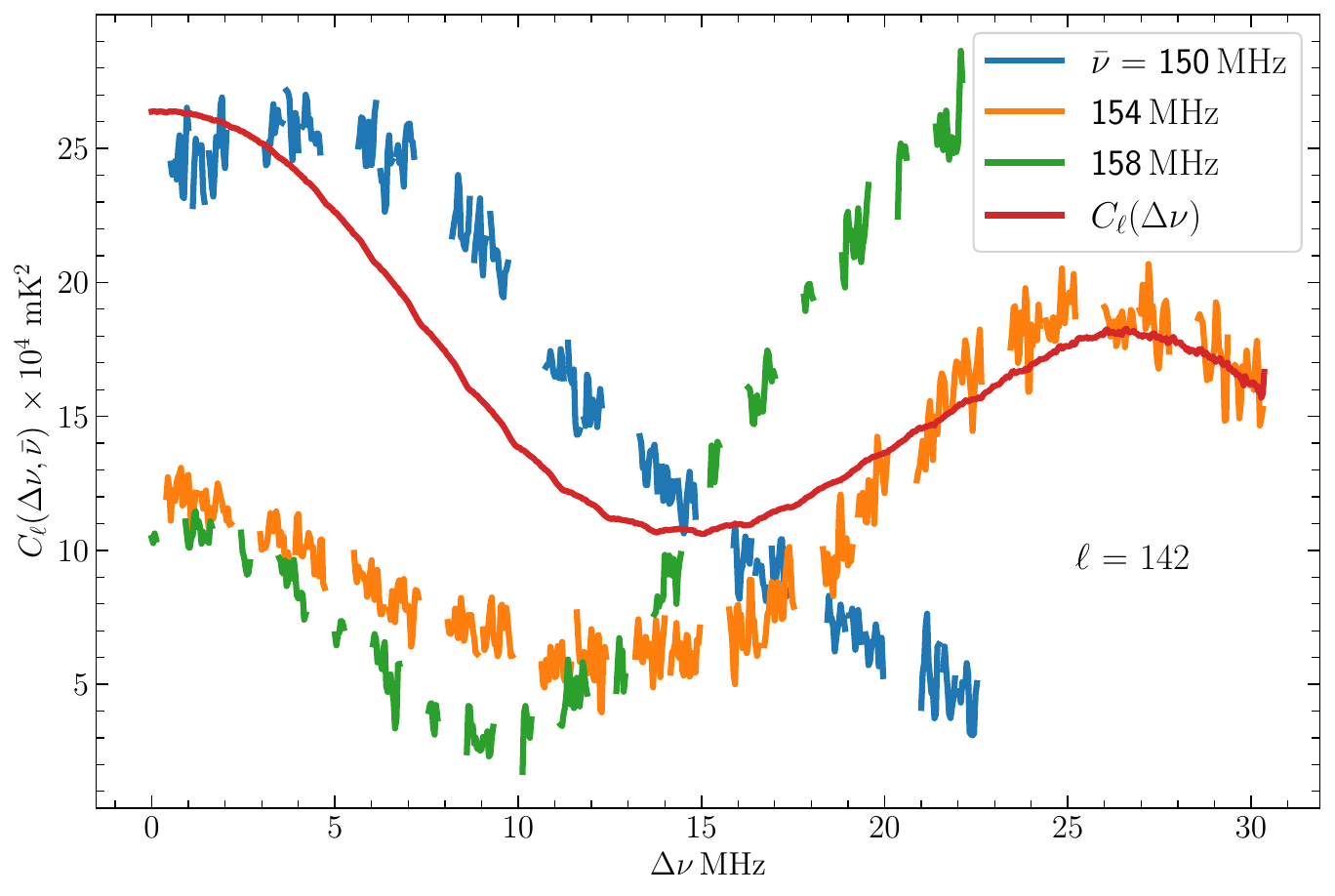}
    \caption{The MAPS $C_{\ell}(\dnu,\nub)$ for some fixed values of $\nub$. The red curve shows $\cl(\dnu)$.}
    \label{fig:nubar_slice_cl}
\end{figure}

\begin{figure}
   \includegraphics[width=\columnwidth]{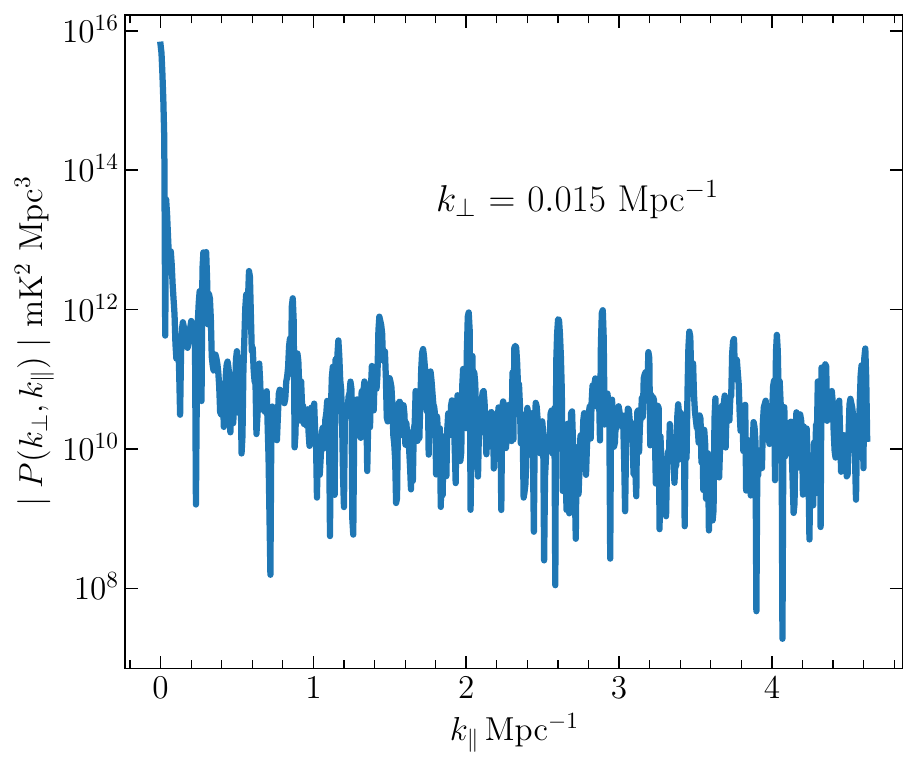}
    \caption{The power spectrum $P(k_\perp, k_\parallel)$ for the MWA data considering the fixed grid $\ell = 142$ for which $\cl(\dnu)$ has been shown in Figure~\ref{fig:nubar_slice_cl} (red curve).}
    \label{fig:pk}
\end{figure}

Figure~\ref{fig:maps_nubar} shows the MAPS $C_{\ell}(\dnu,\nub)$ for the MWA data considering the same grid point mentioned above.  We find that the MAPS is dominated by foregrounds, which are several orders of magnitude larger than the expected 21-cm signal. 
We use this to calculate $C_{\ell}(\dnu)$ by averaging $C_{\ell}(\dnu,\nub)$  along $\nub$, or equivalently collapsing the vertical axis. Note that the missing values are not included in the average. Considering any fixed value of $\dnu$,  we find at least one non-zero value of $C_{\ell}(\dnu,\nub)$ along the vertical direction.  Despite the pattern of missing  $C_{\ell}(\dnu,\nub)$ values, we see in Figure~\ref{fig:nubar_slice_cl}  (red curve) that the resulting  $C_{\ell}(\dnu)$ has no missing values. As a consequence, 
we therefore do not expect the cylindrical PS $\pk$, which is estimated by taking a Fourier transform of  $C_{\ell}(\dnu)$ (eq.~\ref{eq:pk2}),  
to exhibit any artifact due to the missing frequency channels.  However, considering the cylindrical PS $\pk$ shown in Figure~\ref{fig:pk}, 
we see that it shows a regular pattern of spikes at  $\Delta k_{\parallel}=0.29\,{\rm Mpc}^{-1}$. This corresponds to a frequency of $1.28 \, {\rm MHz}$,  
which matches the period of the pattern of flagged channels. Note that the amplitude of these spikes is three to four orders of magnitude smaller than that of the peak foreground value at $k_{\parallel}=0$. These spikes are also considerably smaller than the artifacts that we would obtain due to the missing frequency channels if we were to directly Fourier transform  $\V_{cg}(\nu)$ to obtain  $\V_{cg}(\tau)$ delay space, and use this to estimate $\pk$ (\citealt{Morales2004, Parsons2009, Patwa2021}). Despite being a relatively small effect, these spikes are still several orders of magnitude larger than the expected 21-cm signal. It is necessary to understand the cause of these spikes, and to mitigate their effect  before one can proceed further towards detecting the EoR 21-cm signal. 

We attribute the spikes in $\pk$ (Figure~\ref{fig:pk}) to a tiny ripple that is visible in  $C_{\ell}(\dnu)$ (Figure~\ref{fig:nubar_slice_cl}). The period of this ripple $(\sim 1.28 \, {\rm MHz})$ approximately matches the period of the flagged channels in the MWA data\footnote{The tiny ripples are described in details in Appendix~1 of \citetalias{Chatterjee2024}.}.  The issue now is to identify the cause of this ripple. To address this,  we note that  $C_{\ell}(\dnu,\nub)$,   shown in Figure~\ref{fig:maps_nubar}, exhibits a strong spectral $(\nub)$ dependence. For example,  we see that $C_{\ell}(\dnu,\nub)$ shows an oscillatory pattern of a period of $\sim20$~MHz along $\nub$ if we consider  $\dnu=2 \, {\rm MHz}$ fixed. This is further illustrated in Figure~\ref{fig:nubar_slice_cl} which also shows $C_{\ell}(\dnu,\nub)$ for three different values of $\nub$.  We see that the values of $C_{\ell}(\dnu,\nub)$, and the $\dnu$ dependence, are both quite different for the three values of $\nub$ shown here. Also note the gaps in $C_{\ell}(\dnu,\nub)$, these get filled up when we collapse the vertical $\nub$ axis to obtain  $C_{\ell}(\dnu)$. The point is that we sample a different set of $\nub$ values
for each $\dnu$. This is true even in the absence of missing frequency channels, however this  is further modulated  by the flagging pattern when we have missing frequency channels. However, this would not be an issue, but for the fact that $C_{\ell}(\dnu,\nub)$ exhibits a strong spectral dependence. The resulting $C_{\ell}(\dnu)$  becomes sensitive to the exact combination of $\nub$ values that contribute to any particular $\dnu$. We propose that the ripple seen in  $C_{\ell}(\dnu)$ is due to the sampling of  $\nub$ values, given that $C_{\ell}(\dnu,\nub)$ has a strong spectral dependence. 
 
It is believed that the various astrophysical foregrounds all exhibit a smooth, slowly varying spectral behaviour. However, here we see a strong spectral dependence in $C_{\ell}(\dnu,\nub)$, and it is most likely a consequence of the telescope's chromatic response, possibly baseline migration. 
We propose that it is possible to mitigate the ripple in  $C_{\ell}(\dnu)$, and consequently mitigate the spikes in $\pk$, if we can somehow cut down the level of foreground contamination in the data. We expect the  $C_{\ell}(\dnu,\nub)$ after foreground removal to exhibit a weaker $\nub$ dependence, thereby mitigating the spikes in $\pk$. In Section~\ref{sec:SCF}, we have implemented this idea and demonstrated that it works successfully.

\section{Effect of flagging on simulated MWA data}
\label{sec:simflag}

In this section we use simulated data to study how the flagged channels affect power spectrum estimation. The simulated data gives us control over the properties of the signal, and we can individually study the effect for a few distinct cases.  The simulated data has exactly the same pattern of periodic flagged channels as the actual MWA data. As in the previous section, the entire analysis is restricted to the particular grid point $g$, which corresponds to $\ell = 142$.  In addition to this, Appendix~\ref{sec:sim} presents more realistic simulations that we have carried out to primarily validate Smooth Component Filtering,  which we introduce in Section~\ref{sec:SCF}.

\subsection{21-cm signal}
\label{sub:sig}
In this subsection, we analyze the impact of the missing channel on the expected  21-cm signal. Here, we assume the 21-cm signal to be statistically homogeneous in all three directions. Although the light-cone effect is expected to break this assumption along the LoS direction \citep{Mondal2018}, for simplicity,  we have ignored this here.  The simulated 21-cm signal is, by construction, ergodic along the LoS direction, and we do not expect $C_{\ell}(\dnu,\nub)$ to exhibit any $\nub$ dependence, {\it i.e.},  $C_{\ell}(\dnu,\nub)=C_{\ell}(\dnu)$.

\begin{figure}
    \includegraphics[width=\columnwidth]{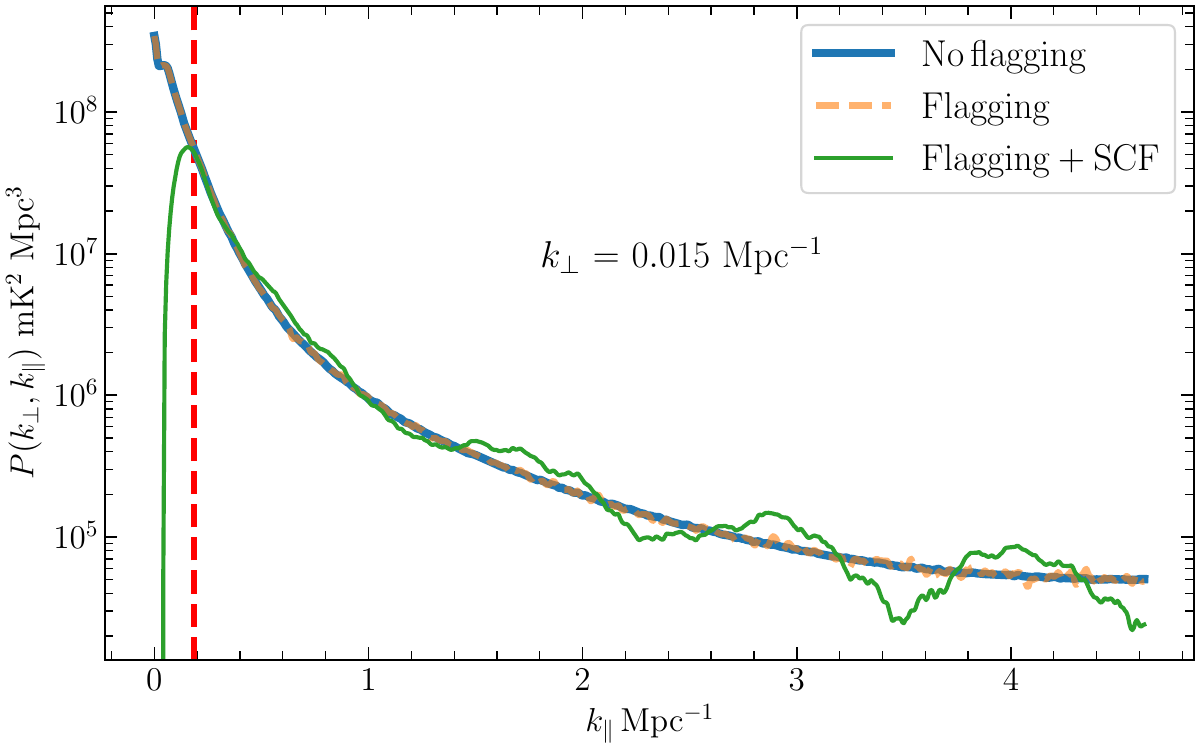}
    \caption{The simulated EoR 21-cm power spectrum $P(k_\perp, k_\parallel)$ as a function of $k_\parallel$ for a fixed value of $k_\perp$. The blue and orange curves, which nearly overlap, show `No flagging' and `Flagging' cases, respectively. The green curve shows the recovered power spectrum after Smooth Component Filtering (SCF) and the red dashed vertical line shows $[k_{\parallel}]_F$, which are  discussed in Section~\ref{sec:SCF}.} 
    \label{fig:21cm_flagging}
\end{figure}

We simulate the expected 21-cm signal  using 
\begin{equation}
      \vcg(\nu)=\int \frac{d k_{\parallel}}{2 \pi} \sqrt{\frac{\pkm}{2 r^2}}  e^{i \, k_{\parallel} r^{'} (\nu-\nu_c)} [x(k_{\parallel})+ i y(k_{\parallel})]
\end{equation}
where $\pkm$ is the input model PS for which we want to simulate the signal, and $x(k_{\parallel})$ and  $y(k_{\parallel})$ are two independent, real   Gaussian random fields that satisfy $\langle x(k_{\parallel}) x(k^{'}_{\parallel}) \rangle = \langle y(k_{\parallel}) y(k^{'}_{\parallel}) \rangle=2 \pi \delta(k_{\parallel}-k^{'}_{\parallel})$.  Here we have used the $z=8$ 21-cm PS from \citet{Mondal2017} as the  the input model PS $\pkm$. Figure~\ref{fig:21cm_flagging} shows  the estimated $\pk$, both with and without flagging.   We see that the two are in close agreement,  and we do not find any artifacts due to the periodic pattern of flagged channels. 

\subsection{Flat spectrum foregrounds}
\label{sub:fsfg}
\begin{figure}
    \includegraphics[width=\columnwidth]{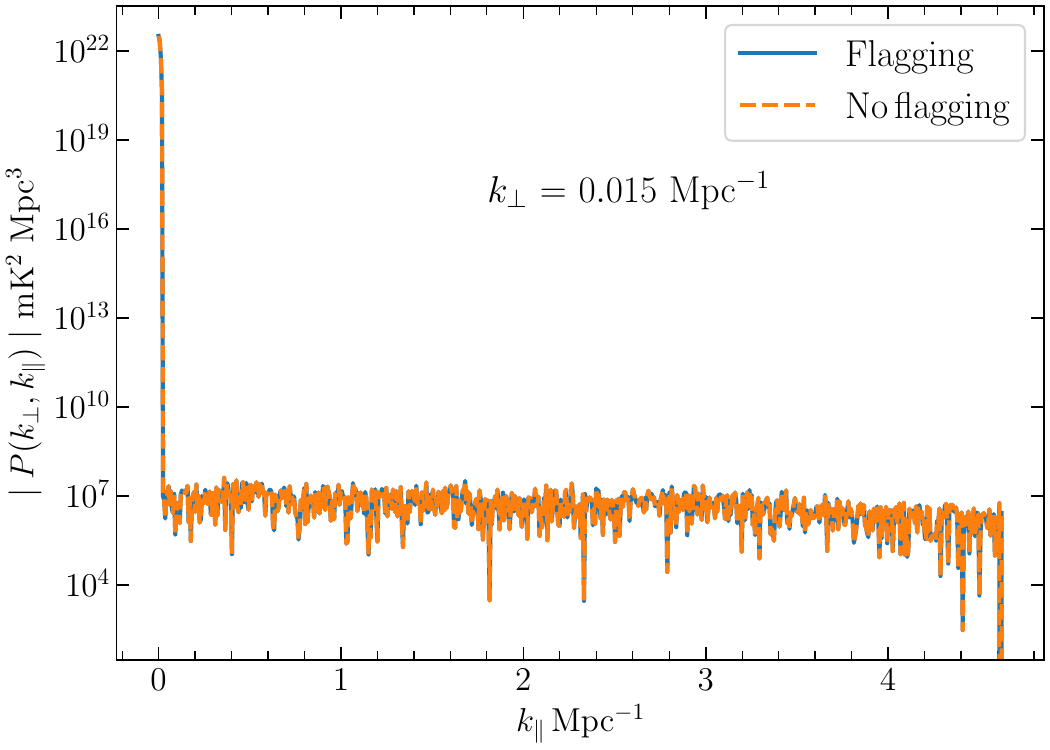}
    \caption{The simulated power spectrum $\mid P(k_\perp, k_\parallel)\mid$ of a flat-spectrum foreground as a function of $k_\parallel$ for a fixed value of $k_\perp$. The blue and orange curves show `Flagging' and `No flagging' cases, respectively. }
    \label{fig:flatfg_flagging}
\end{figure}

In this subsection we consider a flat spectrum foreground signal where both $V_{cg}(\nu)$ and $C_{\ell}(\dnu,\nub)$  are  constant {\it i.e.} have no frequency dependence.  
In this case the missing channels have no impact, and $C_{\ell}(\dnu)$ also is a constant.  We expect $\pk$ to have a  non-zero value only for $k_{\parallel}=0$, with  $\pk=0$ for $k_{\parallel}>0$.  The PS $\pk$ shown in Figure~\ref{fig:flatfg_flagging}   drops sharply by a factor of $\sim 10^{15}$  for $k_{\parallel}>0$. The small non-zero value of $\pk$ at $k_{\parallel}>0$ is a consequence of the limited numerical precision of the computation, and it can be made even smaller be increasing the precision if required. Considering flat spectrum foregrounds, we do not find any artifacts due to the periodic pattern of flagged channels.

\subsection{Power-law spectrum foregrounds.}
\label{sub:plfg}

\begin{figure}
    \includegraphics[width=\columnwidth]{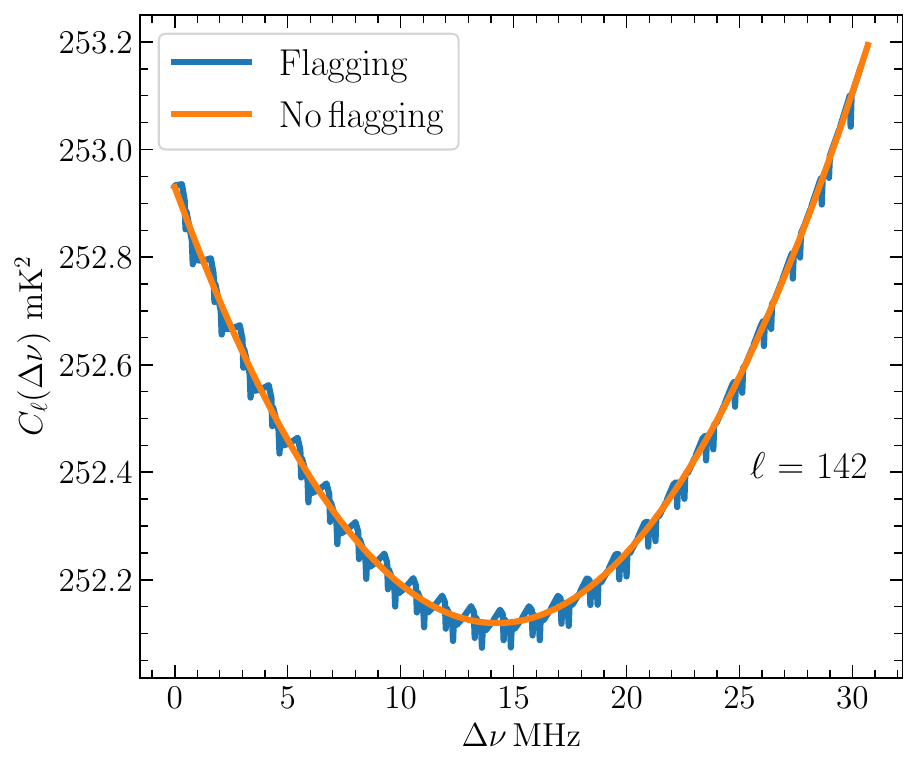}
    \caption{The MAPS $\cl(\dnu)$  for a power law spectrum  foreground $V_{cg} \propto (\nu/\nu_c)^{-\alpha}$, simulated using $\alpha=0.8$. The blue and orange curves show `Flagging' and `No flagging' cases, respectively. }
    \label{fig:cl_specfg}
\end{figure}

\begin{figure}
    \includegraphics[width=\columnwidth]{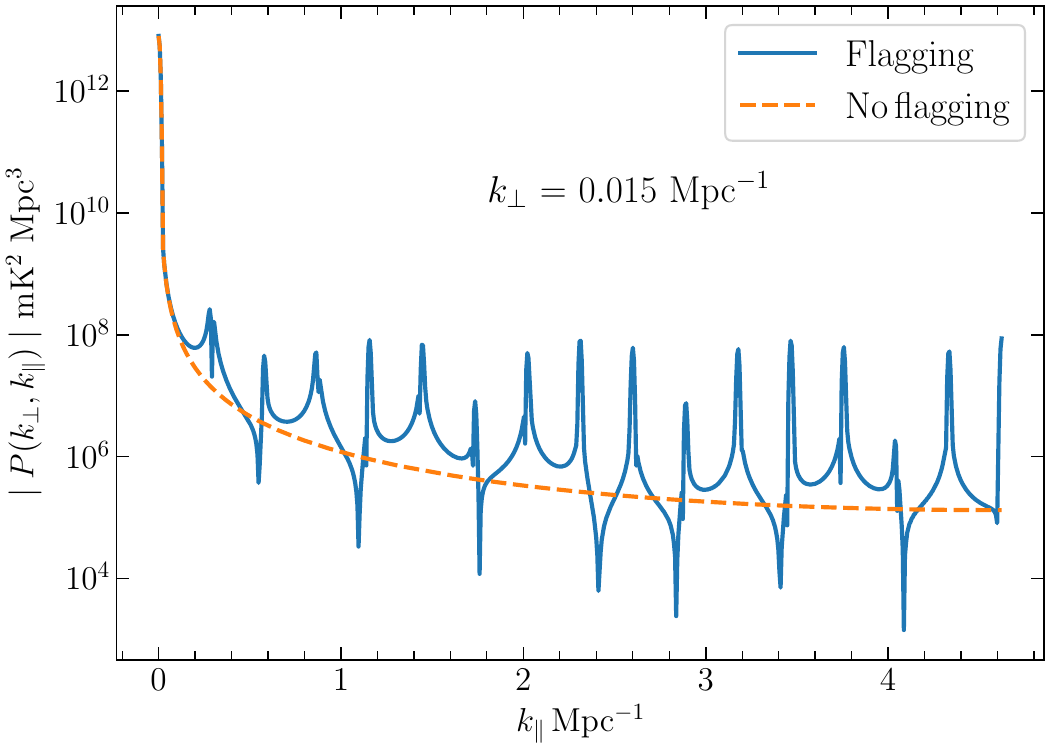}
    \caption{The power spectrum $\mid P(k_\perp, k_\parallel) \mid$ corresponding to the \text{MAPS} $\cl(\dnu)$ shown in Figure~\ref{fig:cl_specfg}. The blue and orange curves show `Flagging' and `No flagging' cases, respectively. }
    \label{fig:specfg_flagging}
\end{figure}

In this subsection we consider a foreground signal that exhibits a power-law spectral dependence  $\V_{cg}(\nu) \propto (\nu/\nu_c)^{-\alpha}$ with $\alpha=0.8$. As discussed in Section~\ref{sec:mwaflag}, we expect the periodic pattern of flagged channels to cause a periodic modulation of the smooth foreground signal. Figure~\ref{fig:cl_specfg} shows $C_{\ell}(\dnu)$ for the simulated signal. In the absence of flagging, $C_{\ell}(\dnu)$ shows a smooth U shaped $\dnu$ dependence, whereas we see a ripple of period $\sim 1.28 \, {\rm MHz}$ superimposed on this when we introduce flagging. Considering $\pk$ shown in   Figure~\ref{fig:specfg_flagging}, we see that the power is mainly concentrated at $k_{\parallel}=0$, and it falls  by a factor in the range  $10^6-10^7$ for $k_{\parallel} \ge  0.4 \, {\rm Mpc}^{-1}$ in the absence of flagging. 
We notice a periodic pattern of spikes, similar to that in  Figure~\ref{fig:pk}, when flagging is introduced. The peak amplitude of the spikes is $\sim 100$ times larger than the base value of $\pk$.  

The results presented here clearly demonstrate that, given the periodic pattern of flagged channels present in the MWA data,  the artifacts in $\pk$ arise due to the strong spectral dependence of the foregrounds. Such artifacts are not present if we consider a statistically homogeneous 21-cm signal or flat spectrum foregrounds.

\section{Smooth Component Filtering (SCF)}
\label{sec:SCF}

We have just seen that the $\pk$ estimated from MWA data  (Figure~\ref{fig:pk}) exhibits artifacts that arise from a combination of two factors, namely the periodic pattern of flagged channels and the strong spectral dependence of the foregrounds. The latter also is possibly introduced by the instrument, and is not an intrinsic property of the foregrounds which are primarily continuum sources that are expected to have a smooth, slowly varying spectral dependence. Given the observational data that carries an imprint of both of these factors, it is not possible to directly eliminate the artifacts from the subsequent PS  estimation. 
Here, we propose that it may be possible to mitigate the artifacts in  $\pk$ if we can somehow reduce the overall amplitude and spectral variation of the foreground contamination in the data. 

The foregrounds, which largely originate from continuum spectra sources,   are expected to have a smoother frequency dependence compared to the \HI{} 21-cm signal that is a line emission.  Here, we have attempted to reduce the overall level of foreground contamination by filtering out the smooth spectral component of $\vcg(\nu)$. We refer to this as {\bf Smooth Component Filtering or SCF}. This will also lead to some loss of the 21-cm signal. As discussed later, we have quantified this and accounted for it in our final results.

We have used a Hann window 
\begin{equation}
    H(n) = \frac{1}{4N} \left[ 1 + \cos \left( 2\pi \frac{n}{2N} \right)   \right] \, , \, \,  -N \leq n \leq N
    \label{eq:Hann}
\end{equation}
to calculate
\begin{equation}
    \vcg^S(\nu_n) = (  \vcg * H )(\nu_n) = \sum_{m=-N}^{N} \vcg(\nu_{m}) H(n-m)\,,
    \label{eq:conv}
\end{equation}
which is the smooth component of $\vcg(\nu)$. Here, the smoothing scale is decided by the width of the Hann window $H(n)$, {\it that is}, $2 N +1$, and we have used $N=50$, which corresponds to a smoothing scale of $2\, {\rm MHz}$. We have subtracted the smooth component $\vcg^S(\nu)$ to obtain 
\begin{equation}
    \vcg^F(\nu) =\vcg(\nu) - \vcg^S(\nu) \,,
    \label{eq:res}
\end{equation}
which is the filtered component that  we  have used  to estimate $C_{\ell}( \nu_a,\nu_b)$  (eq.~\ref{eq:maps3})  analyzed  in the subsequent analysis.  

It is necessary to account for the flagged channels in the actual implementation of the convolution  (eq.~\ref{eq:conv}).  Here, we have used $\vcg(\nu_i) = 0$ for the channels $\nu_i$ that are flagged.  To ensure the correct normalisation after smoothing, we have  divided $\vcg^S(\nu)$ obtained from eq.~(\ref{eq:conv}) with $(U * H)(\nu)$,  where  $U(\nu_i)=0$ for the flagged channels and $U(\nu_i)=1$ otherwise.  The smoothing is restricted to a smaller range  near the boundaries, and we have discarded  $N$ channels  from both ends of $\vcg^F(\nu)$ to account for this. 

We note that we have explored several window functions in addition to the Hann window used here. In addition, we have also tried a range of different smoothing scales for the Hann window. The details of these investigations are presented in Appendix~\ref{sec:window}. Based on this, we have adopted the Hann window with a 2~MHz smoothing scale for the present work.

\begin{figure}
    \includegraphics[width=0.8\columnwidth]{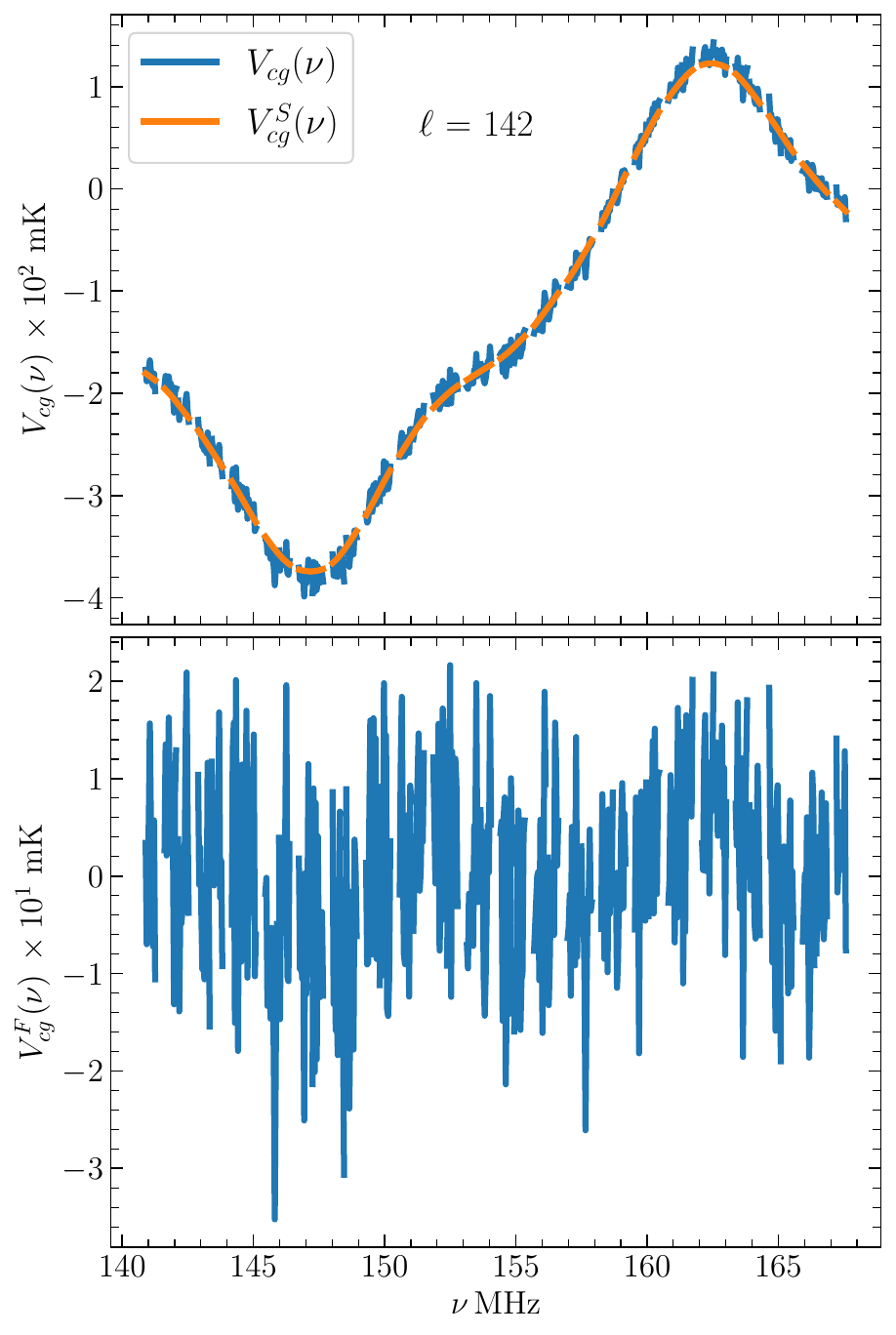}
    \caption{The top panel shows the real part of $\vcg(\nu)$ and $\vcg^S(\nu)$ for the MWA data considering the particular grid point $\ell = 142$. The bottom panel shows $\vcg^F(\nu)$.}
    \label{fig:vcgsmooth}
\end{figure}

The upper panel of Figure~\ref{fig:vcgsmooth} shows the real part of  $\vcg(\nu)$ and $\vcg^S(\nu)$ for the same grid for which the results have been shown in Section~\ref{sec:mwaflag}. We see that $\vcg^S(\nu)$ closely matches $\vcg(\nu)$, and the filtered component  $\vcg^F(\nu)$, shown in the lower panel of  Figure~\ref{fig:vcgsmooth}, is one order of magnitude smaller as compared to $\vcg(\nu)$. We note that the slowly varying, smooth component has been largely subtracted out, and $\vcg^F(\nu)$ primarily shows rapid fluctuation with frequency. 

\begin{figure}
   \includegraphics[width=\columnwidth]{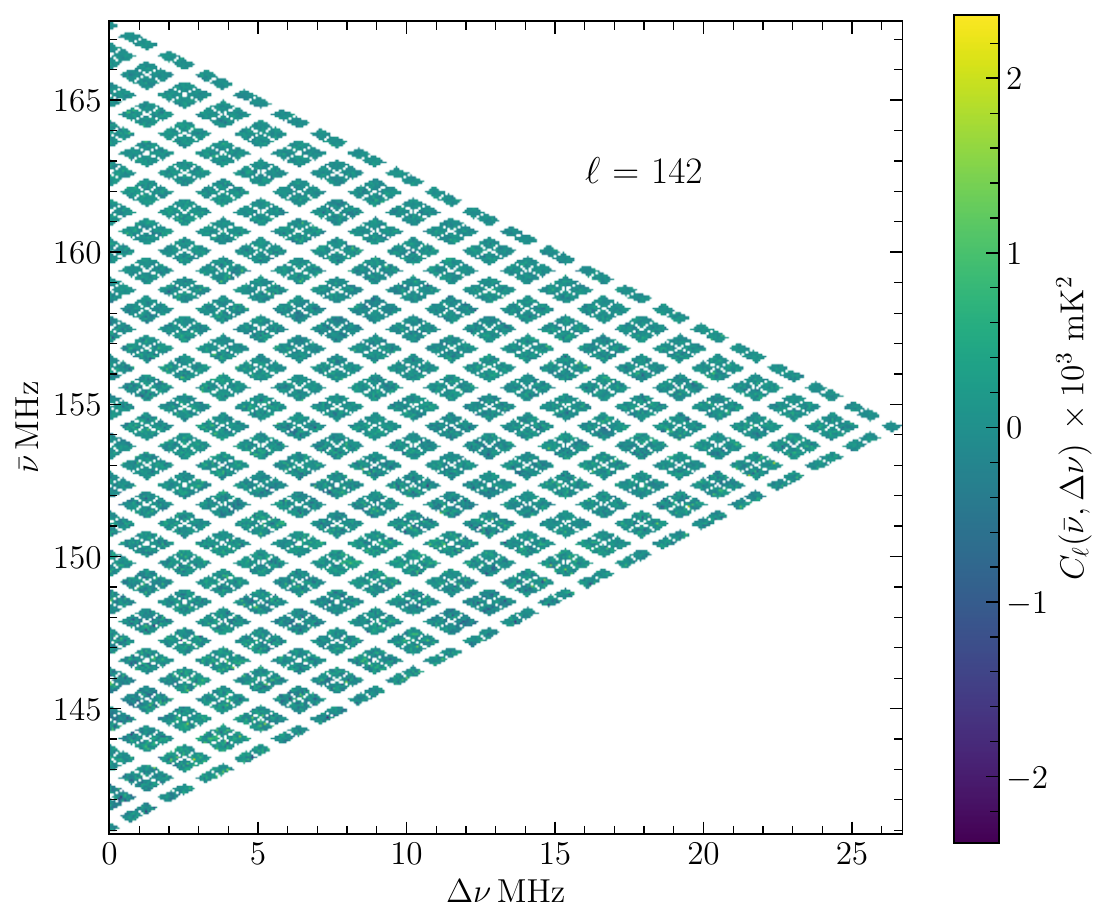}
    \caption{The same as Figure~\ref{fig:maps_nubar}, however after applying SCF.} 
    \label{fig:maps2r-nubar}
\end{figure}

\begin{figure}
   \includegraphics[width=\columnwidth]{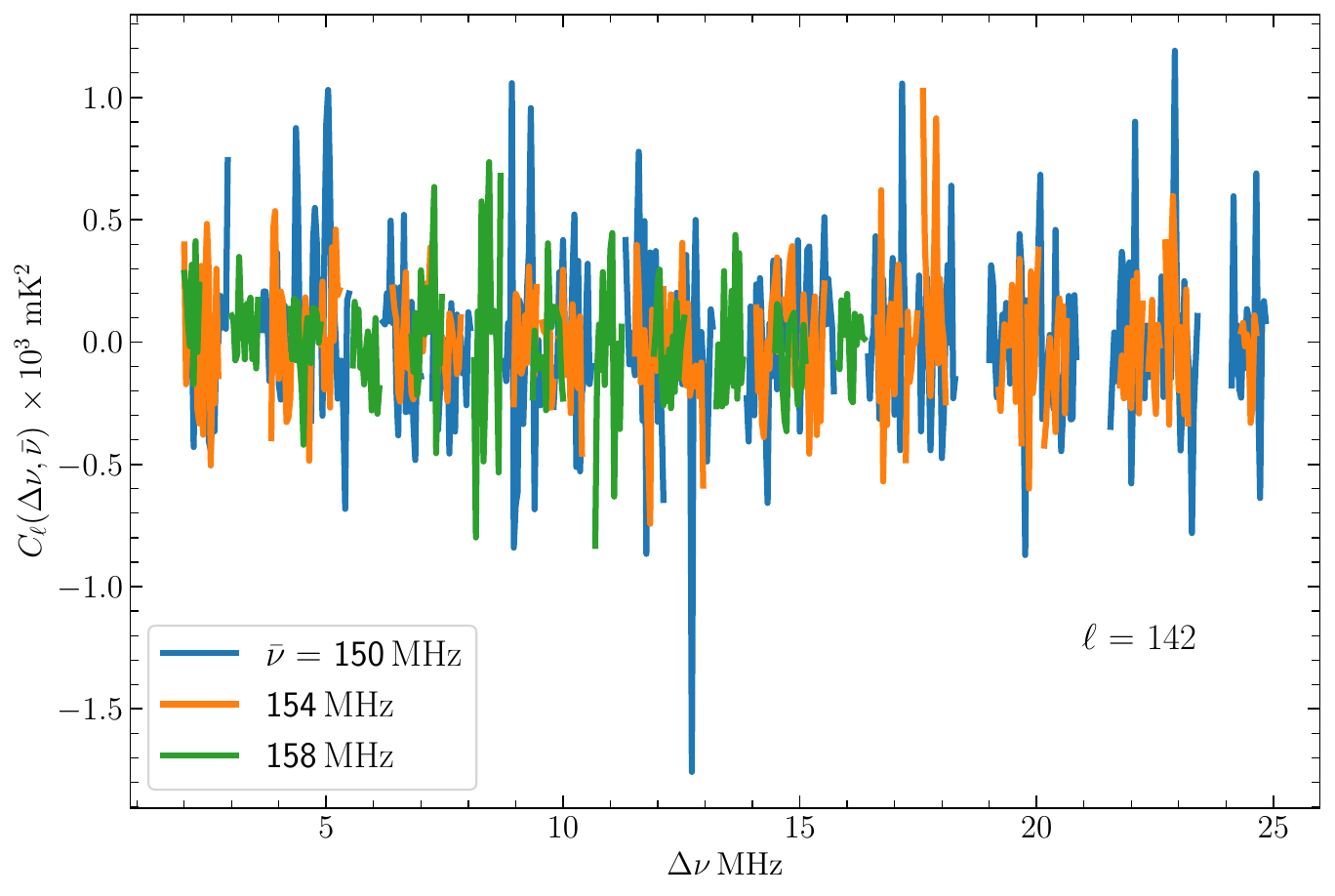}
    \caption{The MAPS $C_{\ell}(\dnu,\nub)$ for the same $\nub$ shown earlier in Figure~\ref{fig:nubar_slice_cl}, however after applying SCF. } 
    \label{fig:maps2r-nubar-slice}
\end{figure}

Figure~\ref{fig:maps2r-nubar} shows  $C_{\ell}(\dnu,\nub)$ that has been calculated using $\vcg^F(\nu)$. As expected, the values have gone down by two orders of magnitude in comparison to those shown in Figure~\ref{fig:maps_nubar}. More importantly, we do not see any strong spectral $(\nub)$ dependence in Figure~\ref{fig:maps2r-nubar}. 
This is further illustrated in Figure~\ref{fig:maps2r-nubar-slice}, which  shows $C_{\ell}(\dnu,\nub)$ as a function of $\dnu$ for the same three values of $\nub$ as in Figure~\ref{fig:nubar_slice_cl}. Comparing the results in the two figures, we see that the values of $C_{\ell}(\dnu,\nub)$  in Figure~\ref{fig:maps2r-nubar-slice} do not show any strong $\nub$ dependence. 

\begin{figure}
   \includegraphics[width=\columnwidth]{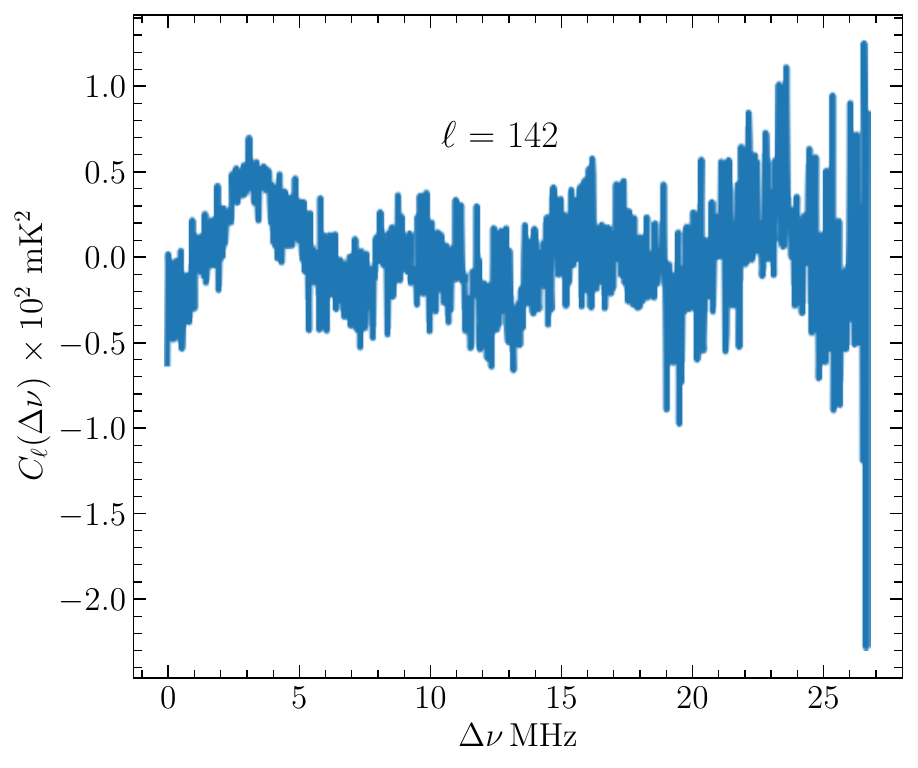}
   \includegraphics[width=\columnwidth]{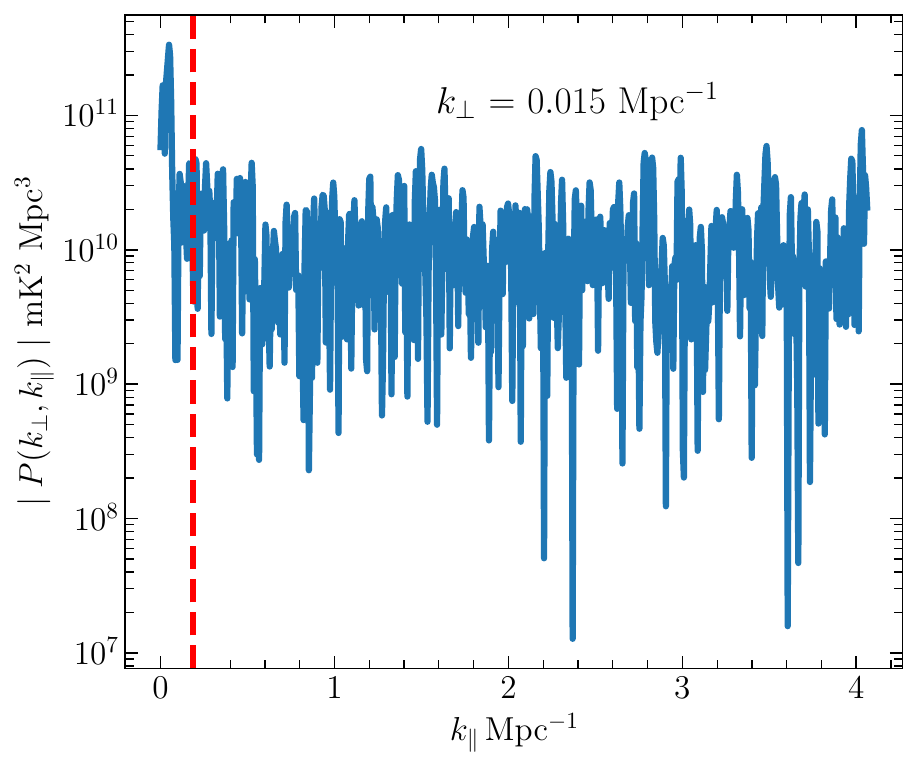}
    \caption{The MAPS $\cl(\dnu)$ (top) and power spectrum $P(k_\perp, k_\parallel)$ (bottom) for the MWA data considering the particular grid ($\ell$ = 142) after SCF. The red dashed vertical line shows $[k_{\parallel}]_F$.}
    \label{fig:pkr}
\end{figure}

The upper panel of Figure~\ref{fig:pkr} shows the $C_{\ell}(\dnu)$ that is obtained after SCF.  Note that we no longer have the smooth $\dnu$ variation that is present in $C_{\ell}(\dnu)$ prior to SCF  (Figure~\ref{fig:maps_nubar}).
We also see that the ripple that  is present in $C_{\ell}(\dnu)$ prior to SCF, is mitigated after SCF.  The lower panel of Figure~\ref{fig:pkr} shows $\pk$ that is obtained after SCF. Comparing this with $\pk$ prior to SCF 
(Figure~\ref{fig:pk}), we see that the periodic pattern of spikes is mitigated after SCF. 

\begin{figure}
   \includegraphics[width=\columnwidth]{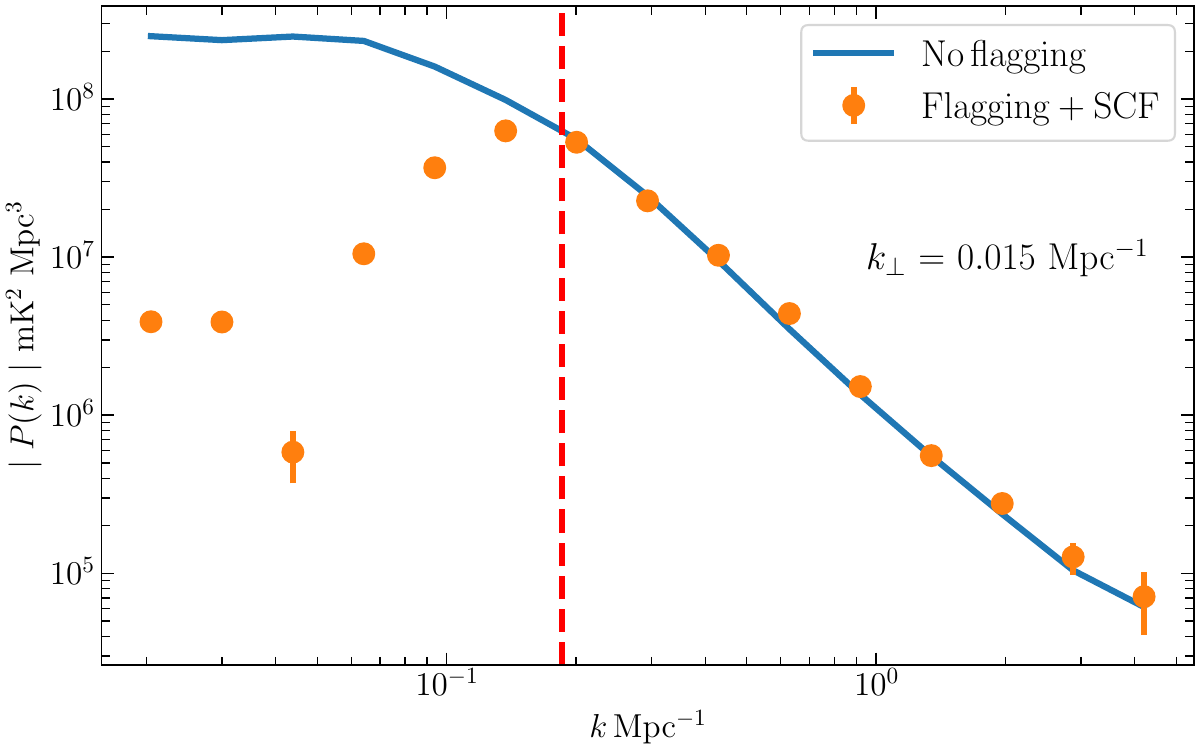}
    \caption{This blue curve shows the spherical power spectrum $P(k)$ for the simulated 21-cm signal considering the particular grid ($\ell$ = 142) with no flagging.  The orange circles show the recovered 21-cm power spectrum after we introduce the periodic flagging and apply SCF. The red dashed vertical line shows $k = \sqrt{ k_\perp ^2 + [k_{\parallel}]_F ^2 }$. }
    \label{fig:21cmSCF}
\end{figure}

SCF is also expected to remove some of the 21-cm signal. To quantify this, we have applied the same filtering  to the simulated 21-cm signal described in Section \ref{sub:sig}.  Figure~\ref{fig:21cm_flagging} shows $\pk$ obtained after applying SCF to the 21-cm signal.  The first point to note is that the power is suppressed at $k_{\parallel}<[k_{\parallel}]_F=2 \pi/(r^{'} \, 2 \, {\rm MHz})=0.185\, {\rm  Mp}c^{-1}$ because we have filtered out the slowly varying component of the signal with a smoothing scale $2 \, {\rm MHz}$.   We are able to recover the expected 21-cm signal at $k_{\parallel} \ge [k_{\parallel}]_F$,  however  $\pk$ exhibits an extra oscillatory feature  along $k_{\parallel}$. This oscillatory feature possibly arises because the smooth signal itself is modulated by the periodic pattern of flagged channels.  Figure~\ref{fig:21cmSCF} shows the same $\pk$, however we have spherically binned the data to present $P(k)$ as a function of $k$. Further, the $k$ bins are of equal logarithmic interval, and are similar to the binning scheme used for the final analysis for which the results are shown in Section~\ref{sec:MWA}. We see that the oscillatory feature averages out within the individual bins, and for $k \ge \sqrt{[k_{\parallel}]^2_F+k^2_{\perp}}$  the estimated values of $P(k)$ are in good agreement  with the $P(k)$ obtained in the absence of flagging and SCF. This is further validated by more realistic simulations presented in Appendix~\ref{sec:sim}.


\section{Results}
\label{sec:MWA}

We have considered baselines in the range  $6 \lambda  \le  U <  220 \lambda$,  where following \citetalias{Chatterjee2024},  we have discarded the baselines $U < 6 \lambda$.  The analysis closely follows \citetalias{Chatterjee2024}, with the difference that we have applied SCF to the $\V_{cg}(\nu)$ for all the grid points, and used  $\V_{cg}^F(\nu)$ to estimate the PS. The baseline range was divided into $20$ equally spaced linear bins, for which we have estimated $C_{\ell}(\dnu)$. These were then used to estimate $\pk$ using the method outlined in Section~\ref{sec:MAPS}, and presented in detail in \citetalias{Chatterjee2024}.

\begin{figure*}
    \includegraphics[width=\textwidth]{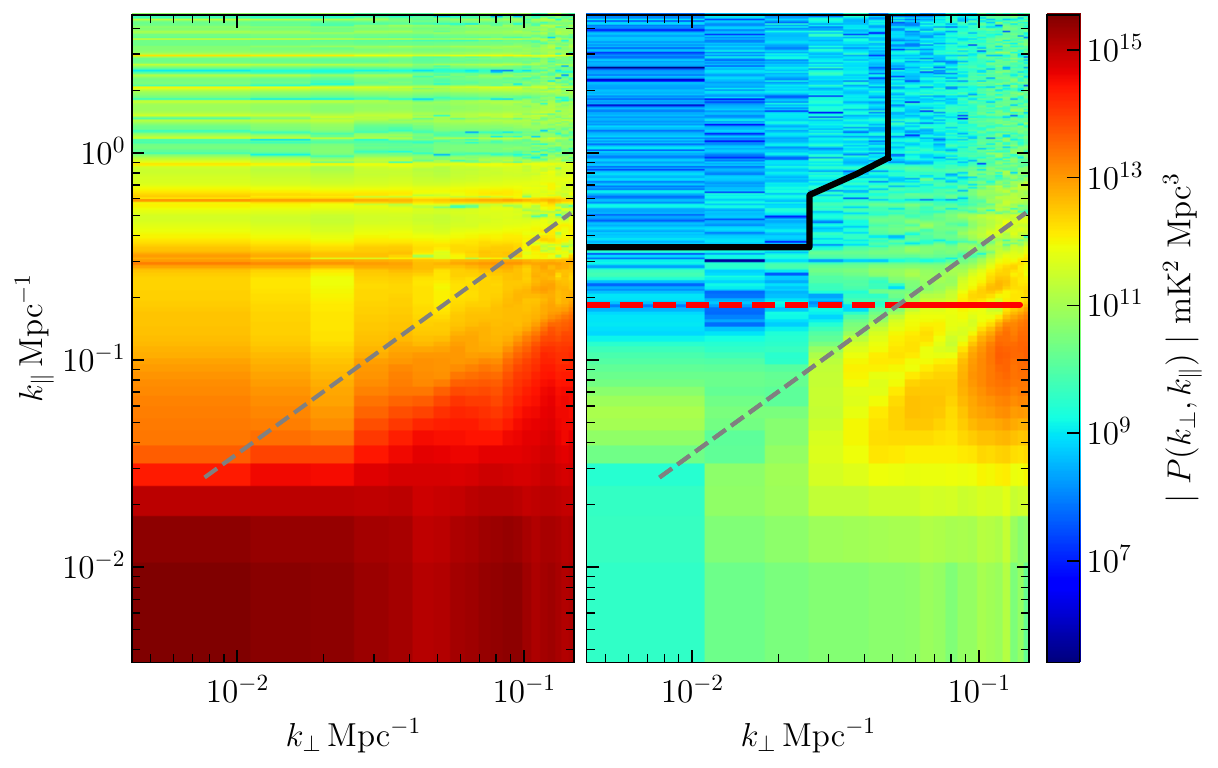}
    \caption{The cylindrical PS $\mid P(\kpp, \kpar) \mid$ before (left) and after (right) SCF. The grey dashed curve in both panels shows the theoretically predicted boundary of the foreground wedge. The red dashed curve shows $[k_{\parallel}]_F=0.185\, {\rm  Mp}c^{-1}$, below which  $(k_{\parallel}<[k_{\parallel}]_F)$  SCF filters out the signal. The black  lines demarcate the region that we have used to constrain the 21-cm PS, the $(\kpp, \kpar)$ modes are restricted to the range     $\kpar > 0.35  \, {\rm Mpc^{-1}}$ and     $\kpp  \le 0.048 \, {\rm Mpc^{-1}}$.} 
    \label{fig:cylps}
\end{figure*}

The right panel of Figure~\ref{fig:cylps} shows $\pk$  estimated after applying SCF to the gridded visibilities $\V_{cg}(\nu)$, while the left panel shows the same when SCF is not applied. Considering the left panel, we see a very pronounced foreground wedge, with some leakage extending beyond the wedge. We also note a periodic pattern of horizontal streaks that spans nearly the entire $(\kpp, k_{\parallel})$ plane. These streaks correspond to the spikes that appear (Figure \ref{fig:pk}) due to the periodic pattern of flagged channels. These streaks contaminate most of the EoR-window, and it was possible to use only a very small region of the $(\kpp,k_{\parallel})$ plane  to constrain the 21-cm PS in \citetalias{Chatterjee2024}. We expect SCF to remove most of the power at $k_{\parallel}<[k_{\parallel}]_F=0.185\, {\rm  Mp}c^{-1}$, and we cannot use this region to constrain the 21-cm PS. Considering the right panel, we see that the level of foreground contamination is significantly reduced at $k_{\parallel}  > [k_{\parallel}]_F$ for the short baselines  ($\kpp  \le 0.5 \, {\rm Mpc}^{-1}$), however, there is still some visible foreground contamination at the long baselines ($\kpp  > 0.5 \, {\rm Mpc}^{-1}$) even after we apply SCF.  
This arises due to baseline migration that causes the foreground signal to oscillate rapidly with frequency at long baselines. As a consequence, for long baselines, the foreground wedge extends to $k_{\parallel}>[k_{\parallel}]_F$. This is supported by realistic foreground simulations presented in Appendix~\ref{sec:sim}.  Our analysis indicates that we have to use a smoothing scale smaller than $2 \, {\rm MHz}$. It may be possible to overcome this limitation by using a smaller smoothing scale at the longer baselines, but we have not considered this here.  We have discarded the large $\kpp$, and only used the range $0.004 \leq \kpp \leq 0.048 \, {\rm Mpc^{-1}}$ for the subsequent analysis. Considering the range  $k_{\parallel} \ge [k_{\parallel}]_F$,  we see that the streaks are considerably mitigated when SCF is applied.   We do not expect SCF to cause any signal loss in this $k_{\parallel}$ range. We have incorporated a small buffer in $\kpar$, and used the range $\kpar > 0.35  \, {\rm Mpc^{-1}}$ to constrain the 21-cm PS. The region of the $(\kpp, k_{\parallel})$ plane that we have used to constrain the 21-cm  PS is indicated in the right panel of Figure~\ref{fig:cylps}. 

\begin{figure}
    \includegraphics[width=\columnwidth]{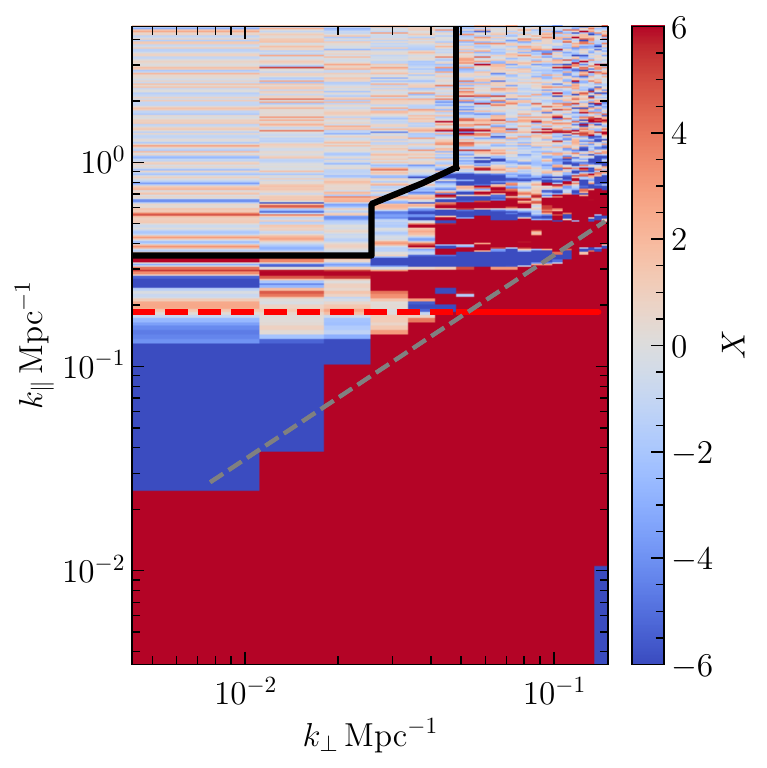}
    \caption{The heat map of $X$ with the color bar being saturated within $\mid X \mid < 6$. The region inside the black curve is used for estimating the spherical PS, and all data points have values $\mid X \mid < 6.79$ in this region.}
    \label{fig:X_hm}
\end{figure}

We assess the statistics of the  estimated $\pk$ values using  the quantity $X$ defined as \citep{Pal2020}
\begin{equation}
    X=\frac{P(\kpp, \kpar)}{\delta P_{N}(\kpp, \kpar)} \, ,
    \label{eq:xstat}
\end{equation}
where $\delta P_{N}(\kpp, \kpar)$  is the statistical uncertainty expected from the system noise only. Figure~\ref{fig:X_hm} shows the values of $X$ for all  the $(\kpp, \kpar)$ modes.  For the modes with $k_{\parallel} \le [k_{\parallel}]_F$, the values of $\mid X \mid$ are found to be very large $(\sim10^5)$  and we have saturated the color bar within $\mid X \mid < 6$ for a better visualization. For the subsequent analysis, we have discarded the range  $k_{\parallel} \le [k_{\parallel}]_F$  where we expect a large signal loss due to the  smoothing.  Considering the range  $k_{\parallel} \ge [k_{\parallel}]_F$,  we see that the values of $\mid X \mid$ are considerably smaller for the short baselines  ($\kpp  \le 0.5 \, {\rm Mpc}^{-1}$) in comparison to the long baselines ($\kpp  >  0.5 \, {\rm Mpc}^{-1}$) which are still contaminated by the residual foregrounds that are not filtered by SCF. We have discarded the long baselines  $(\kpp  > 0.5 \, {\rm Mpc}^{-1})$ and  we only use the first 6 $\kpp$ bins in the subsequent analysis. To avoid additional foreground leakage, we have used the range $\kpar > 0.35  \, {\rm Mpc}^{-1}$ for the first three  $\kpp$ bins, while we have used a larger buffer $6\times k_{\rm H}$  for the next three $\kpp$ bins, where $k_{\rm H}$ is the theoretically predicted boundary of the foreground wedge. 
The $(\kpp, \kpar)$ region that we have used to estimate the final EoR 21-cm PS is demarcated using a black solid curve throughout this paper. For this selected region, the maximum value of $\mid X \mid$ is 6.79, and $99.63\%$ values are within $\mid X \mid \le 5$.

\begin{figure}
    \includegraphics[width=\columnwidth]{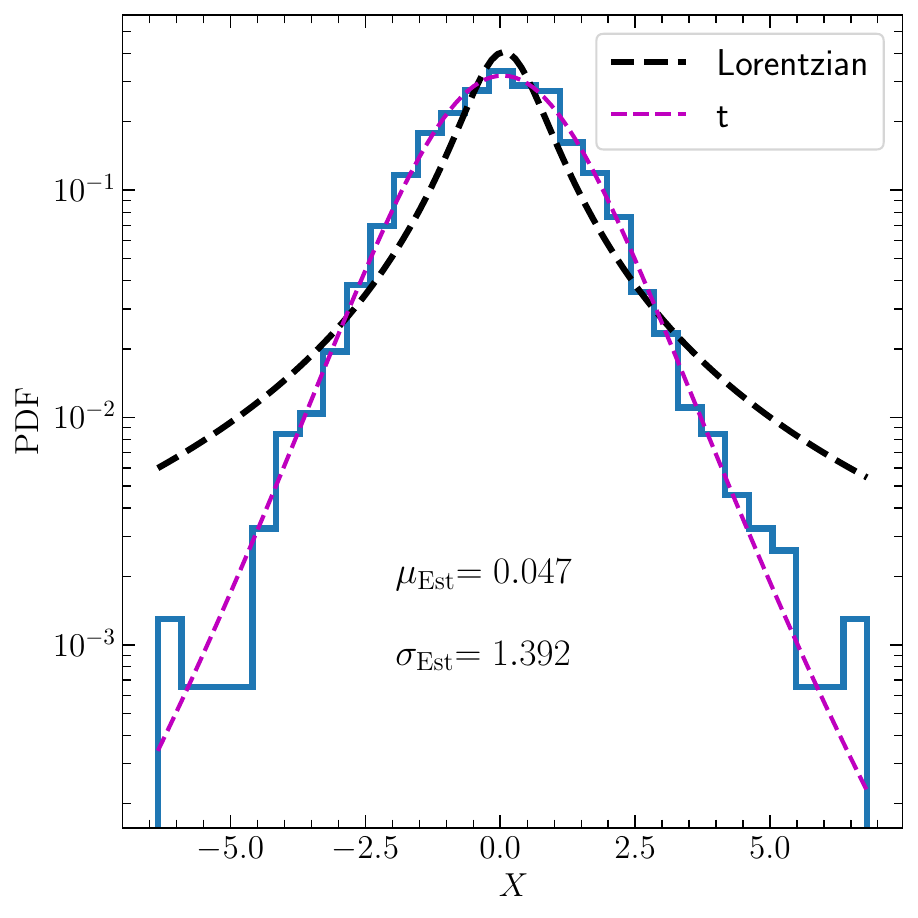}
    \caption{The histogram shows the probability density function (PDF) of $X$. The two dashed curves show the best-fit  Lorentzian (black) and  t-distribution (magenta) respectively.}
    \label{fig:X}
\end{figure}

As detailed in \citetalias{Chatterjee2024} (and also in \citealt{Pal2022, Elahi2023, Elahi2023b, Elahi2024}), we expect $X$ to have a symmetric distribution with mean $\mu = 0$ and standard deviation $\sigma = 1$ if the estimated $\pk$ values are consistent with the uncertainties predicted due to the system noise contribution only. We expect any residual foreground contamination in $\pk$ to make the distribution of $X$ asymmetric towards a positive value and produce a positive mean, whereas any negative systematics introduced in the analysis are expected to have the opposite effect.  
Figure~\ref{fig:X} shows the estimated probability density function (PDF) of $X$, which we see is nearly symmetric with $\mu_{\rm Est}=0.047$ and  $\sigma_{\rm Est}=1.392$. Applying the Student's t-test, we find that the null hypothesis that the estimated distribution of $X$ is consistent with zero mean has a $p$-value of 0.04, which is acceptable. We therefore conclude that the estimated $\pk$ values that are used for the subsequent analysis do not show any evidence of residual foregrounds or negative systematics.   However, the value $\sigma_{\rm Est}=1.392$ indicates that the measured $P(\kpp, \kpar)$ has fluctuations that are larger than those predicted from the system noise only. A possible origin of this `excess variance' can be traced to the strong $\nub$ dependence of $C_{\ell}(\dnu,\nub)$. Considering Figure~\ref{fig:nubar_slice_cl} that does not incorporate  SCF, we see a that the strong $\nub$ dependence is expected to introduce large fluctuations in $C_{\ell}(\dnu)$, much larger than those expected from the system noise alone. Note that we have $\sigma_{\rm Est} \approx 95$  \citepalias{Chatterjee2024} when we do not incorporate SCF. Applying SCF reduces the $\nub$ dependence (Figure~\ref{fig:maps2r-nubar-slice}), and the level of fluctuations is reduced. However, there is still some residual $\nub$ dependence which manifests itself as excess fluctuations, and we have   $\sigma_{\rm Est}=1.392$. We note that several earlier works \citep{Mertens2020, Pal2020} and  \citep{Pal2022, Elahi2023, Elahi2023b, Elahi2024} have reported such excess variance for the EoR and post-EoR  21-cm power spectrum respectively. 
We note that, the value of $\sigma_{\rm Est}$ increases if we include the modes from $\kpp  > 0.5 \, {\rm Mpc}^{-1}$ as the residual foreground contaminations are higher in these $\kpp$  bins. On the other hand the value of $\sigma_{\rm Est}$ changes between $1.2-1.4$ if we discard some of the $\kpp$  bins which are within $\kpp \le 0.5 \, {\rm Mpc}^{-1}$. To achieve the best constraints on the 21-cm signal, we have used as many $\kpp$ bins as possible, as long as the distribution of $X$ is symmetric.
We further note that the PDF of $X$ is well-modelled by a t-distribution, which is similar to the findings of \cite{Elahi2023, Elahi2023b, Elahi2024}.  Following \cite{Pal2020, Pal2022, Elahi2023, Elahi2023b, Elahi2024} in the subsequent analysis, we have scaled the system noise only error predictions with $\sigma_{\rm Est}$ to account for the excess variance {\it i.e.} we have used $ \delta P(\kpp, \kpar) =  \sigma_{\rm Est} \times   \delta P_{N}(\kpp, \kpar)  $ to predict the statistical fluctuations expected in the measured $P(\kpp, \kpar)$.

\begin{figure}
    \includegraphics[width=\columnwidth]{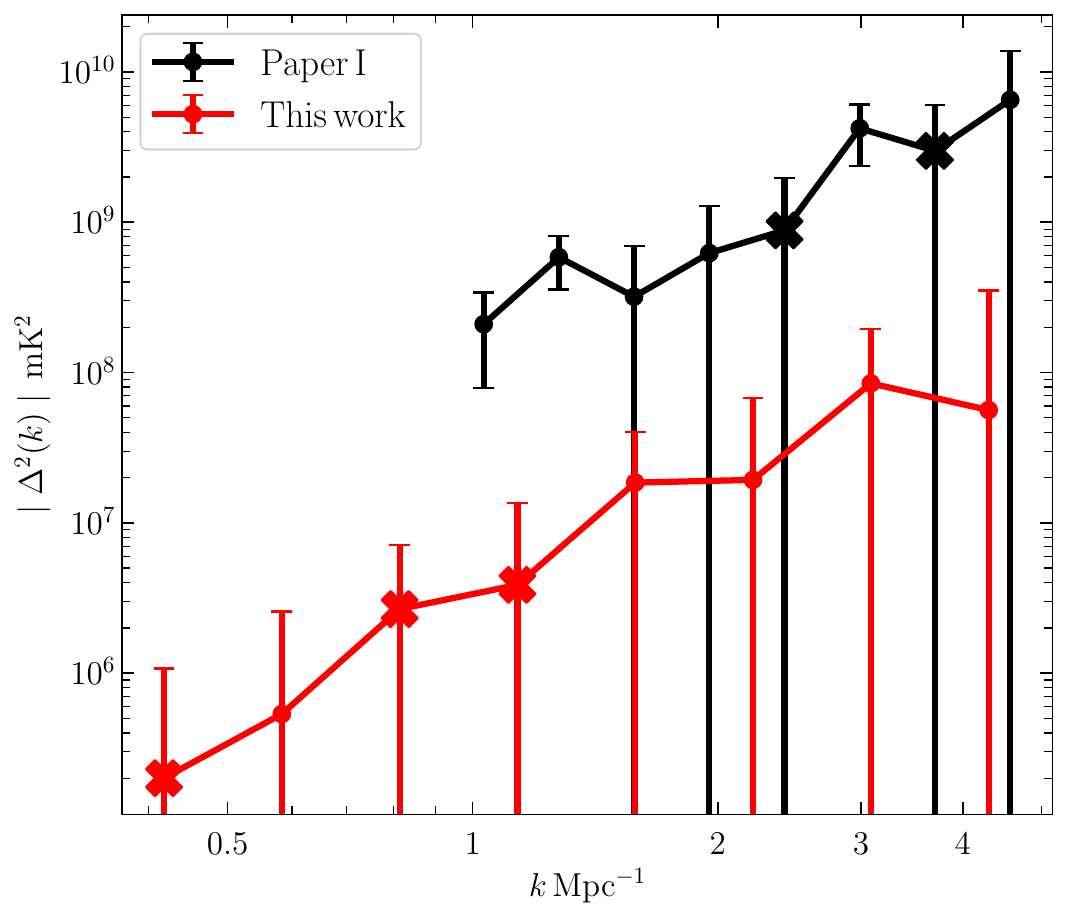}
    \caption{This red curve shows the  measured $\mid \Delta^2(k) \mid $ and  $2\sigma$ uncertainties. The results from \citetalias{Chatterjee2024} are shown in black for reference. Negative values of $\Delta^2(k)$ are indicated using cross (x) marks.
  }
    \label{fig:sphericalps}
\end{figure}

\begin{table}
    \centering
    \caption{The measured $\Delta^2(k)$, corresponding errors $\sigma(k)$, ${\rm SNR} = \Delta^2(k)/\sigma(k)$, and the $2\sigma$ upper limits $\Delta_{\rm UL}^{2}(k)$.  }
        \begin{tabular}{ccccc}
        \hline
        \hline
        $k$ & $\Delta^2(k)$ & $\sigma(k)$ & SNR & $\Delta_{\rm UL}^{2}(k)$  \\
        $\rm{Mpc}^{-1}$ & $\rm{mK}^2$ & $\rm{mK}^2$ & & $\rm{mK}^2$\\
        \hline
        $0.418$ &  $-(447.46)^2$ & $(660.86)^2$ & $-0.46$ & $(934.60)^2$\\
        $0.583$ &  $(731.46)^2$ & $(1009.24)^2$ & $0.53$ & $(1603.80)^2$\\
        $0.813$ &  $-(1633.44)^2$ & $(1499.97)^2$ & $-1.19$ & $(2121.28)^2$\\
        $1.135$ &  $-(1966.73)^2$ & $(2190.81)^2$ & $-0.81$ & $(3098.28)^2$\\
        $1.584$ &  $(4304.10)^2$ & $(3301.66)^2$ & $1.70$ & $(6350.37)^2$\\
        $2.210$ &  $(4395.35)^2$ & $(4941.38)^2$ & $0.79$ & $(8255.52)^2$\\
        $3.085$ &  $(9203.94)^2$ & $(7467.56)^2$ & $1.52$ & $(14008.62)^2$\\
        $4.305$ &  $(7500.37)^2$ & $(12154.35)^2$ & $0.38$ & $(18753.99)^2$\\
        \hline
    \end{tabular}
    \label{tab:ul_MLE}
\end{table}

We now consider the spherical power spectrum $P(k)$. Following \citet{Elahi2023}, we have estimated this directly from $C_{\ell}(\dnu)$ considering $8$ bins of uniform logarithmic interval and using  only the $(\kpp, \kpar)$ region shown in Figure~\ref{fig:cylps}. 
Figure~\ref{fig:sphericalps} shows the mean squared brightness temperature fluctuations $\Delta^2(k) = k^3 P(k)/(2 \pi^2)$ as a function of $k$ along with the $2\sigma$ error bars corresponding to the predicted statistical fluctuations $\delta P(\kpp, \kpar)$. The values are tabulated in Table~\ref{tab:ul_MLE}, which also presents  $\Delta_{\rm UL}^{2}(k)$ the $2\sigma$ upper limits of $\Delta^2(k)$. The results from \citetalias{Chatterjee2024} are also shown in 
Figure~\ref{fig:sphericalps} for comparison. The results in \citetalias{Chatterjee2024} span the $k$ range  $k=1 - 4 \, {\rm Mpc}^{-1} $, and we find the tightest upper limit of $\Delta_{\rm UL}^2(k) =(1.85\times10^4)^2\, {\rm mK^2}$ at the first $k$-bin $k=1 \,{\rm Mpc}^{-1}$.
Considering the present work, we find that the values of $\Delta^2(k)$ are around $\sim 50$ times smaller. The $k$ range extends to smaller values, the values of $\Delta_{\rm UL}^2(k)$ in the first seven $k$ bins are consistent with $0 \pm 2\sigma$  and we now have the tightest upper limit of $\Delta_{\rm UL}^2(k) =(934.60)^2 \, {\rm mK^2}$ at the first $k$-bin $k=0.418 \,{\rm Mpc}^{-1}$. We find that the tightest upper limit $\Delta_{\rm UL}^2(k)$ is improved by a factor of $\sim 400$ after we apply SCF, which is a major improvement for 21-cm PS estimation.
Note, however, that the upper limits obtained here are still several orders of magnitude larger than the values in the range $\Delta^2(k) \sim 10^1-10^2 \, {\rm mK}^2 $ predicted for the  EoR  21-cm signal across the $k$  range considered here \citep{Mondal2017}. 

\section{Summary and Conclusions}
\label{sec:summary}

We have analyzed how the periodic pattern of flagged channels affects 21-cm PS estimation from MWA data.  We particularly focus on the MAPS $C_{\ell}(\dnu)$, and the cylindrical PS $\pk$ which is calculated from $C_{\ell}(\dnu)$ through a Fourier transform along $\dnu$. We find that the flagged channels do not cause any missing $\dnu$ in $C_{\ell}(\dnu)$. However, the missing channels introduce a ripple in $C_{\ell}(\dnu)$, whose period matches that of the flagged channels. This ripple introduces a periodic pattern of spikes along $\kpar$ in $\pk$.

In order to trace the cause of the ripple, we have analyzed the MAPS $C_{\ell}(\dnu,\nub)$, where the extra $\nub$ dependence allows us to trace the spectral evolution of the signal within the frequency band of observation. We find that the measured  $C_{\ell}(\dnu,\nub)$, which is foreground dominated, shows significant spectral ($\nub$) dependence. Using simulations, we demonstrate that the ripple in $C_{\ell}(\dnu)$ arises from a combination of the periodic patterns of flagged channels and the strong spectral dependence of the foregrounds. The simulations show that flagged channels do not introduce any artifacts in $C_{\ell}(\dnu)$ or $\pk$ for the statistically homogeneous 21-cm signal or for spectrally flat foregrounds, for both of which $C_{\ell}(\dnu,\nub)$ does not exhibit any $\nub$ dependence. 

Here we have addressed the problem by introducing smooth component filtering (SCF) that removes the slowly varying spectral component of the measured visibility data $\V_{cg}(\nu)$. The filtering has been implemented by smoothing $\V_{cg}(\nu)$ on $2 \, {\rm MHz}$ scale, and subtracting out this smooth component. We find that the $C_{\ell}(\dnu,\nub)$ estimated from the filtered visibility data $\V_{cg}^F(\nu)$  does not exhibit the strong $\nub$ dependence that is present prior to SCF.  We find that the ripple in $C_{\ell}(\dnu)$, and the spikes in $\pk$ are considerably mitigated after SCF. The filtering leads to considerable signal loss at low $\kpar$, which we exclude from the subsequent analysis. We also find that the filtering is not very effective at large $\kpp$ where baseline migration causes the foreground signal to oscillate at a scale faster than $2 \, {\rm MHz}$. It may be possible to overcome this by reducing the smoothing scale at longer baselines ($\kpp$).  We have not considered this here, and we have excluded the large $\kpp$ from the subsequent analysis.  However,  we  still find a considerably large region of the $ (\kpp, \kpar)$  plane (Figure~\ref{fig:cylps})
that can be used to constrain the 21-cm PS reliably. 

We see that it is possible to mitigate the spikes in $\pk$ by using SCF, which filters out the dominant, spectrally smooth foreground component from the visibility data. An alternate, and possibly superior approach would be to accurately model the foregrounds from an image and subtract these from the visibility data. One possibility is to use a direction-dependent calibration along with spectral regularization to simultaneously model the smooth foregrounds and calibrate the instrument  (e.g. \citealt{Yatawatta2024, Brackenhoff2025}). Such an approach typically requires observations with a high angular resolution and noise sensitivity. The present observation, which is a shallow zenith pointing snapshot, has very limited baseline coverage and noise sensitivity. It offers very limited scope for imaging to reliably identify the foregrounds, and we have not attempted this here.

Our results for $\Delta^2(k)$ the mean-squared 21-cm brightness temperature fluctuations are shown in Figure~\ref{fig:sphericalps}, and summarized in Table~\ref{tab:ul_MLE}. We obtain the tightest $2\sigma$ upper limit of  $\Delta_{\rm UL}^2(k) =(934.60)^2 \, {\rm mK^2}$ at the smallest  $k$-bin $k=0.418 \,{\rm Mpc}^{-1}$. This is a factor of $\sim 400$ improvement on our earlier results \citepalias{Chatterjee2024} that did not incorporate SCF. However, the upper limit is still several orders of magnitude larger than the predicted EoR  21-cm signal
$(\Delta^2(k) \sim 10^1-10^2 \, {\rm mK}^2)$.
The results presented here correspond to a single pointing from a drift-scan observation that covers 162 different pointings on the sky \citep{Patwa2021}. 
The results 
are found to be within $0 \pm 2\sigma$, {\it i.e.}, they are consistent with statistical fluctuations, and it should be possible to improve the upper limits by combining observations in other pointing directions \citep{Chatterjee2022}. In future work, we plan to analyze the data corresponding to the other pointing directions and combine these to target tighter constrains of the 21-cm PS.

\section*{Acknowledgements}
We thank the anonymous reviewer for the comments and suggestions which helped us to improve the paper.
A. Elahi thanks Thiagaraj Prabu for useful discussions. A. Elahi acknowledges the computing facilities of IIT Kharagpur and IIT Madras.
S. Chatterjee acknowledges support from the South African National Research Foundation (Grant No. 84156) and the Inter-University Institute for Data Intensive Astronomy (IDIA). IDIA is a partnership of the University of Cape Town, the University of Pretoria and the University of the Western Cape. S. Chatterjee would also like to thank Dr. Devojyoti Kansabanik, for helpful discussions.
We acknowledge the use of the ilifu cloud computing facility – www.ilifu.ac.za, a partnership between the University of Cape Town, the University of the Western Cape, Stellenbosch University, Sol Plaatje University and the Cape Peninsula University of Technology. The ilifu facility is supported by contributions from the Inter-University Institute for Data Intensive Astronomy (IDIA – a partnership between the University of Cape Town, the University of Pretoria and the University of the Western Cape), the Computational Biology division at UCT and the Data Intensive Research Initiative of South Africa (DIRISA).

\section*{Data Availability} 
The data sets were derived from sources in the public domain
(the MWA Data Archive: project ID G0031) at \url{https://asvo.mwatelescope.org/}.



\bibliographystyle{mnras}
\bibliography{mylist} 

\begin{thebibliography}{}
\makeatletter
\relax
\def\mn@urlcharsother{\let\do\@makeother \do\$\do\&\do\#\do\^\do\_\do\%\do\~}
\def\mn@doi{\begingroup\mn@urlcharsother \@ifnextchar [ {\mn@doi@} {\mn@doi@[]}}
\def\mn@doi@[#1]#2{\def\@tempa{#1}\ifx\@tempa\@empty \href {http://dx.doi.org/#2} {doi:#2}\else \href {http://dx.doi.org/#2} {#1}\fi \endgroup}
\def\mn@eprint#1#2{\mn@eprint@#1:#2::\@nil}
\def\mn@eprint@arXiv#1{\href {http://arxiv.org/abs/#1} {{\tt arXiv:#1}}}
\def\mn@eprint@dblp#1{\href {http://dblp.uni-trier.de/rec/bibtex/#1.xml} {dblp:#1}}
\def\mn@eprint@#1:#2:#3:#4\@nil{\def\@tempa {#1}\def\@tempb {#2}\def\@tempc {#3}\ifx \@tempc \@empty \let \@tempc \@tempb \let \@tempb \@tempa \fi \ifx \@tempb \@empty \def\@tempb {arXiv}\fi \@ifundefined {mn@eprint@\@tempb}{\@tempb:\@tempc}{\expandafter \expandafter \csname mn@eprint@\@tempb\endcsname \expandafter{\@tempc}}}

\bibitem[\protect\citeauthoryear{{Abdurashidova} et~al.,}{{Abdurashidova} et~al.}{2022}]{Abdurashidova2022}
{Abdurashidova} Z.,  et~al., 2022, \mn@doi [\apj] {10.3847/1538-4357/ac1c78}, \href {https://ui.adsabs.harvard.edu/abs/2022ApJ...925..221A} {925, 221}

\bibitem[\protect\citeauthoryear{{Acharya}, {Mertens}, {Ciardi}, {Ghara}, {Koopmans}  \& {Zaroubi}}{{Acharya} et~al.}{2024}]{Acharya2024}
{Acharya} A.,  {Mertens} F.,  {Ciardi} B.,  {Ghara} R.,  {Koopmans} L. V.~E.,   {Zaroubi} S.,  2024, \mn@doi [\mnras] {10.1093/mnrasl/slae078}, \href {https://ui.adsabs.harvard.edu/abs/2024MNRAS.534L..30A} {534, L30}

\bibitem[\protect\citeauthoryear{{Ali}, {Bharadwaj}  \& {Chengalur}}{{Ali} et~al.}{2008}]{Ali2008}
{Ali} S.~S.,  {Bharadwaj} S.,   {Chengalur} J.~N.,  2008, \mn@doi [\mnras] {10.1111/j.1365-2966.2008.12984.x}, \href {http://adsabs.harvard.edu/abs/2008MNRAS.385.2166A} {385, 2166}

\bibitem[\protect\citeauthoryear{{Bandura} et~al.,}{{Bandura} et~al.}{2014}]{Bandura2014}
{Bandura} K.,  et~al., 2014, in Ground-based and Airborne Telescopes V. p. 914522 (\mn@eprint {arXiv} {1406.2288}), \mn@doi{10.1117/12.2054950}

\bibitem[\protect\citeauthoryear{{Barry}, {Hazelton}, {Sullivan}, {Morales}  \& {Pober}}{{Barry} et~al.}{2016}]{Barry16}
{Barry} N.,  {Hazelton} B.,  {Sullivan} I.,  {Morales} M.~F.,   {Pober} J.~C.,  2016, \mn@doi [\mnras] {10.1093/mnras/stw1380}, \href {https://ui.adsabs.harvard.edu/abs/2016MNRAS.461.3135B} {461, 3135}

\bibitem[\protect\citeauthoryear{{Barry}, {Beardsley}, {Byrne}, {Hazelton}, {Morales}, {Pober}  \& {Sullivan}}{{Barry} et~al.}{2019}]{Barry2019eppsilon}
{Barry} N.,  {Beardsley} A.~P.,  {Byrne} R.,  {Hazelton} B.,  {Morales} M.~F.,  {Pober} J.~C.,   {Sullivan} I.,  2019, \mn@doi [\pasa] {10.1017/pasa.2019.21}, \href {https://ui.adsabs.harvard.edu/abs/2019PASA...36...26B} {36, e026}

\bibitem[\protect\citeauthoryear{{Bernardi} et~al.,}{{Bernardi} et~al.}{2009}]{Bernardi2009}
{Bernardi} G.,  et~al., 2009, \mn@doi [\aap] {10.1051/0004-6361/200911627}, \href {http://adsabs.harvard.edu/abs/2009A%26A...500..965B} {500, 965}

\bibitem[\protect\citeauthoryear{{Bharadwaj} \& {Ali}}{{Bharadwaj} \& {Ali}}{2005}]{Bharadwaj2005}
{Bharadwaj} S.,  {Ali} S.~S.,  2005, \mn@doi [\mnras] {10.1111/j.1365-2966.2004.08604.x}, \href {http://adsabs.harvard.edu/abs/2005MNRAS.356.1519B} {356, 1519}

\bibitem[\protect\citeauthoryear{{Bharadwaj} \& {Sethi}}{{Bharadwaj} \& {Sethi}}{2001}]{Bharadwaj2001b}
{Bharadwaj} S.,  {Sethi} S.~K.,  2001, \mn@doi [J. Astrophys. Astron.] {10.1007/BF02702273}, \href {http://adsabs.harvard.edu/abs/2001JApA...22..293B} {22, 293}

\bibitem[\protect\citeauthoryear{{Bharadwaj}, {Nath}  \& {Sethi}}{{Bharadwaj} et~al.}{2001}]{Bharadwaj2001a}
{Bharadwaj} S.,  {Nath} B.~B.,   {Sethi} S.~K.,  2001, \mn@doi [J. Astrophys. Astron.] {10.1007/BF02933588}, \href {http://adsabs.harvard.edu/abs/2001JApA...22...21B} {22, 21}

\bibitem[\protect\citeauthoryear{Bharadwaj, Pal, Choudhuri  \& Dutta}{Bharadwaj et~al.}{2018}]{Bharadwaj2018}
Bharadwaj S.,  Pal S.,  Choudhuri S.,   Dutta P.,  2018, \mn@doi [\mnras] {10.1093/mnras/sty3501}, 483, 5694

\bibitem[\protect\citeauthoryear{Blake, Ferreira  \& Borrill}{Blake et~al.}{2004}]{Blake2004b}
Blake C.,  Ferreira P.~G.,   Borrill J.,  2004, \mn@doi [\mnras] {10.1111/j.1365-2966.2004.07831.x}, 351, 923

\bibitem[\protect\citeauthoryear{{Brackenhoff} et~al.,}{{Brackenhoff} et~al.}{2025}]{Brackenhoff2025}
{Brackenhoff} S.~A.,  et~al., 2025, \mn@doi [arXiv e-prints] {10.48550/arXiv.2504.02483}, \href {https://ui.adsabs.harvard.edu/abs/2025arXiv250402483B} {p. arXiv:2504.02483}

\bibitem[\protect\citeauthoryear{{Byrne} et~al.,}{{Byrne} et~al.}{2019}]{Byrne2019}
{Byrne} R.,  et~al., 2019, \mn@doi [\apj] {10.3847/1538-4357/ab107d}, \href {https://ui.adsabs.harvard.edu/abs/2019ApJ...875...70B} {875, 70}

\bibitem[\protect\citeauthoryear{{Byrne}, {Morales}, {Hazelton}, {Sullivan}, {Barry}, {Lynch}, {Line}  \& {Jacobs}}{{Byrne} et~al.}{2022}]{Byrne2022}
{Byrne} R.,  {Morales} M.~F.,  {Hazelton} B.,  {Sullivan} I.,  {Barry} N.,  {Lynch} C.,  {Line} J. L.~B.,   {Jacobs} D.~C.,  2022, \mn@doi [\mnras] {10.1093/mnras/stab3276}, \href {https://ui.adsabs.harvard.edu/abs/2022MNRAS.510.2011B} {510, 2011}

\bibitem[\protect\citeauthoryear{{CHIME Collaboration} et~al.,}{{CHIME Collaboration} et~al.}{2022}]{Amiri2022}
{CHIME Collaboration} et~al., 2022, \mn@doi [\apjs] {10.3847/1538-4365/ac6fd9}, \href {https://ui.adsabs.harvard.edu/abs/2022ApJS..261...29C} {261, 29}

\bibitem[\protect\citeauthoryear{{Ceccotti} et~al.,}{{Ceccotti} et~al.}{2025}]{Ceccotti2025}
{Ceccotti} E.,  et~al., 2025, \mn@doi [\aap] {10.1051/0004-6361/202453106}, \href {https://ui.adsabs.harvard.edu/abs/2025A&A...696A..56C} {696, A56}

\bibitem[\protect\citeauthoryear{Chakraborty et~al.,}{Chakraborty et~al.}{2021}]{Chakraborty2021}
Chakraborty A.,  et~al., 2021, \mn@doi [The Astrophysical Journal Letters] {10.3847/2041-8213/abd17a}, 907, L7

\bibitem[\protect\citeauthoryear{Chakraborty, Datta  \& Mazumder}{Chakraborty et~al.}{2022}]{Chakraborty2022}
Chakraborty A.,  Datta A.,   Mazumder A.,  2022, \mn@doi [The Astrophysical Journal] {10.3847/1538-4357/ac5cc5}, 929, 104

\bibitem[\protect\citeauthoryear{{Chapman} et~al.,}{{Chapman} et~al.}{2012}]{Chapman2012}
{Chapman} E.,  et~al., 2012, \mn@doi [\mnras] {10.1111/j.1365-2966.2012.21065.x}, \href {http://adsabs.harvard.edu/abs/2012MNRAS.423.2518C} {423, 2518}

\bibitem[\protect\citeauthoryear{Chatterjee, Bharadwaj  \& Marthi}{Chatterjee et~al.}{2020}]{Chatterjee2021}
Chatterjee S.,  Bharadwaj S.,   Marthi V.~R.,  2020, \mn@doi [\mnras] {10.1093/mnras/staa3348}, 500, 4398

\bibitem[\protect\citeauthoryear{Chatterjee, Bharadwaj, Choudhuri, Sethi  \& Patwa}{Chatterjee et~al.}{2022}]{Chatterjee2022}
Chatterjee S.,  Bharadwaj S.,  Choudhuri S.,  Sethi S.,   Patwa A.~K.,  2022, \mn@doi [\mnras] {10.1093/mnras/stac3576}, 519, 2410

\bibitem[\protect\citeauthoryear{{Chatterjee}, {Elahi}, {Bharadwaj}, {Sarkar}, {Choudhuri}, {Sethi}  \& {Patwa}}{{Chatterjee} et~al.}{2024}]{Chatterjee2024}
{Chatterjee} S.,  {Elahi} K. M.~A.,  {Bharadwaj} S.,  {Sarkar} S.,  {Choudhuri} S.,  {Sethi} S.~K.,   {Patwa} A.~K.,  2024, \mn@doi [\pasa] {10.1017/pasa.2024.45}, \href {https://ui.adsabs.harvard.edu/abs/2024PASA...41...77C} {41, e077}

\bibitem[\protect\citeauthoryear{{Chatterjee}, {Sarkar}, {Choudhuri}, {Elahi}, {Bharadwaj}, {Sethi}  \& {Patwa}}{{Chatterjee} et~al.}{2025}]{Chatterjee2025}
{Chatterjee} S.,  {Sarkar} S.,  {Choudhuri} S.,  {Elahi} K. M.~A.,  {Bharadwaj} S.,  {Sethi} S.~K.,   {Patwa} A.~K.,  2025, \pasa

\bibitem[\protect\citeauthoryear{{Chen} et~al.,}{{Chen} et~al.}{2025}]{Chen2025}
{Chen} K.-F.,  et~al., 2025, \mn@doi [\apj] {10.3847/1538-4357/ad9b91}, \href {https://ui.adsabs.harvard.edu/abs/2025ApJ...979..191C} {979, 191}

\bibitem[\protect\citeauthoryear{{Choudhuri}, {Bharadwaj}, {Ghosh}  \& {Ali}}{{Choudhuri} et~al.}{2014}]{Choudhuri2014}
{Choudhuri} S.,  {Bharadwaj} S.,  {Ghosh} A.,   {Ali} S.~S.,  2014, \mn@doi [\mnras] {10.1093/mnras/stu2027}, \href {http://adsabs.harvard.edu/abs/2014MNRAS.445.4351C} {445, 4351}

\bibitem[\protect\citeauthoryear{Choudhuri, Bharadwaj, Chatterjee, Ali, Roy  \& Ghosh}{Choudhuri et~al.}{2016}]{Choudhuri2016b}
Choudhuri S.,  Bharadwaj S.,  Chatterjee S.,  Ali S.~S.,  Roy N.,   Ghosh A.,  2016, \mn@doi [\mnras] {10.1093/mnras/stw2254}, 463, 4093

\bibitem[\protect\citeauthoryear{Choudhuri, Bharadwaj, Ali, Roy, Intema  \& Ghosh}{Choudhuri et~al.}{2017}]{Choudhuri2017}
Choudhuri S.,  Bharadwaj S.,  Ali S.~S.,  Roy N.,  Intema H.~T.,   Ghosh A.,  2017, \mn@doi [\mnras: Letters] {10.1093/mnrasl/slx066}, 470, L11

\bibitem[\protect\citeauthoryear{{Datta}, {Choudhury}  \& {Bharadwaj}}{{Datta} et~al.}{2007}]{Datta2007}
{Datta} K.~K.,  {Choudhury} T.~R.,   {Bharadwaj} S.,  2007, \mn@doi [\mnras] {10.1111/j.1365-2966.2007.11747.x}, \href {http://adsabs.harvard.edu/abs/2007MNRAS.378..119D} {378, 119}

\bibitem[\protect\citeauthoryear{{Datta}, {Bowman}  \& {Carilli}}{{Datta} et~al.}{2010}]{Datta2010}
{Datta} A.,  {Bowman} J.~D.,   {Carilli} C.~L.,  2010, \mn@doi [\apj] {10.1088/0004-637X/724/1/526}, \href {http://adsabs.harvard.edu/abs/2010ApJ...724..526D} {724, 526}

\bibitem[\protect\citeauthoryear{De~Oliveira-Costa, Tegmark, Gaensler, Jonas, Landecker  \& Reich}{De~Oliveira-Costa et~al.}{2008}]{DeOliveiraCosta2008}
De~Oliveira-Costa A.,  Tegmark M.,  Gaensler B.~M.,  Jonas J.,  Landecker T.~L.,   Reich P.,  2008, \mn@doi [\mnras] {10.1111/j.1365-2966.2008.13376.x}, 388, 247

\bibitem[\protect\citeauthoryear{DeBoer et~al.,}{DeBoer et~al.}{2017}]{Deboer2017}
DeBoer D.~R.,  et~al., 2017, Publications of the Astronomical Society of the Pacific, 129, 045001

\bibitem[\protect\citeauthoryear{{Dillon}}{{Dillon}}{2015}]{Dillon2015b}
{Dillon} J.~S.,  2015, preprint, \href {http://adsabs.harvard.edu/abs/2015arXiv150603024D} {} (\mn@eprint {arXiv} {1506.03024})

\bibitem[\protect\citeauthoryear{Dillon et~al.,}{Dillon et~al.}{2014}]{Dillon2014}
Dillon J.~S.,  et~al., 2014, \mn@doi [Phys. Rev. D] {10.1103/PhysRevD.89.023002}, 89, 023002

\bibitem[\protect\citeauthoryear{{Elahi} et~al.,}{{Elahi} et~al.}{2023a}]{Elahi2023}
{Elahi} K. M.~A.,  et~al., 2023a, \mn@doi [\mnras] {10.1093/mnras/stad191}, \href {https://ui.adsabs.harvard.edu/abs/2023MNRAS.520.2094E} {520, 2094}

\bibitem[\protect\citeauthoryear{{Elahi} et~al.,}{{Elahi} et~al.}{2023b}]{Elahi2023b}
{Elahi} K. M.~A.,  et~al., 2023b, \mn@doi [\mnras] {10.1093/mnras/stad2495}, \href {https://ui.adsabs.harvard.edu/abs/2023MNRAS.525.3439E} {525, 3439}

\bibitem[\protect\citeauthoryear{{Elahi} et~al.,}{{Elahi} et~al.}{2024}]{Elahi2024}
{Elahi} K. M.~A.,  et~al., 2024, \mn@doi [\mnras] {10.1093/mnras/stae740}, \href {https://ui.adsabs.harvard.edu/abs/2024MNRAS.529.3372E} {529, 3372}

\bibitem[\protect\citeauthoryear{{Ewall-Wice} et~al.,}{{Ewall-Wice} et~al.}{2021}]{Ewall-Wice2021}
{Ewall-Wice} A.,  et~al., 2021, \mn@doi [\mnras] {10.1093/mnras/staa3293}, \href {https://ui.adsabs.harvard.edu/abs/2021MNRAS.500.5195E} {500, 5195}

\bibitem[\protect\citeauthoryear{Franzen, Vernstrom, Jackson, Hurley-Walker, Ekers, Heald, Seymour  \& White}{Franzen et~al.}{2019}]{Franzen2019}
Franzen T. M.~O.,  Vernstrom T.,  Jackson C.~A.,  Hurley-Walker N.,  Ekers R.~D.,  Heald G.,  Seymour N.,   White S.~V.,  2019, \mn@doi [Publications of the Astronomical Society of Australia] {10.1017/pasa.2018.52}, 36, e004

\bibitem[\protect\citeauthoryear{{Gan} et~al.,}{{Gan} et~al.}{2023}]{Gan2023}
{Gan} H.,  et~al., 2023, \mn@doi [\aap] {10.1051/0004-6361/202244316}, \href {https://ui.adsabs.harvard.edu/abs/2023A&A...669A..20G} {669, A20}

\bibitem[\protect\citeauthoryear{{Gayen}, {Kumar}, {Dutta}, {Elahi}, {Choudhuri}  \& {Roy}}{{Gayen} et~al.}{2025}]{Gayen2025}
{Gayen} S.,  {Kumar} J.,  {Dutta} P.,  {Elahi} K. M.~A.,  {Choudhuri} S.,   {Roy} N.,  2025, \mn@doi [arXiv e-prints] {10.48550/arXiv.2503.23825}, \href {https://ui.adsabs.harvard.edu/abs/2025arXiv250323825G} {p. arXiv:2503.23825}

\bibitem[\protect\citeauthoryear{{Gehlot} et~al.,}{{Gehlot} et~al.}{2022}]{Gehlot2022dgse}
{Gehlot} B.~K.,  et~al., 2022, \mn@doi [\aap] {10.1051/0004-6361/202142939}, \href {https://ui.adsabs.harvard.edu/abs/2022A&A...662A..97G} {662, A97}

\bibitem[\protect\citeauthoryear{{Ghosh}, {Bharadwaj}, {Ali}  \& {Chengalur}}{{Ghosh} et~al.}{2011a}]{Ghosh2011a}
{Ghosh} A.,  {Bharadwaj} S.,  {Ali} S.~S.,   {Chengalur} J.~N.,  2011a, \mn@doi [\mnras] {10.1111/j.1365-2966.2010.17853.x}, \href {http://adsabs.harvard.edu/abs/2011MNRAS.411.2426G} {411, 2426}

\bibitem[\protect\citeauthoryear{{Ghosh}, {Bharadwaj}, {Ali}  \& {Chengalur}}{{Ghosh} et~al.}{2011b}]{Ghosh2011b}
{Ghosh} A.,  {Bharadwaj} S.,  {Ali} S.~S.,   {Chengalur} J.~N.,  2011b, \mn@doi [\mnras] {10.1111/j.1365-2966.2011.19649.x}, \href {http://adsabs.harvard.edu/abs/2011MNRAS.418.2584G} {418, 2584}

\bibitem[\protect\citeauthoryear{{Ghosh}, {Prasad}, {Bharadwaj}, {Ali}  \& {Chengalur}}{{Ghosh} et~al.}{2012}]{Ghosh2012}
{Ghosh} A.,  {Prasad} J.,  {Bharadwaj} S.,  {Ali} S.~S.,   {Chengalur} J.~N.,  2012, \mn@doi [\mnras] {10.1111/j.1365-2966.2012.21889.x}, \href {http://adsabs.harvard.edu/abs/2012MNRAS.426.3295G} {426, 3295}

\bibitem[\protect\citeauthoryear{Gorski, Hivon, Banday, Wandelt, Hansen, Reinecke  \& Bartelmann}{Gorski et~al.}{2005}]{Gorski2005}
Gorski K.~M.,  Hivon E.,  Banday A.~J.,  Wandelt B.~D.,  Hansen F.~K.,  Reinecke M.,   Bartelmann M.,  2005, \mn@doi [\apj] {10.1086/427976}, 622, 759

\bibitem[\protect\citeauthoryear{Gupta et~al.,}{Gupta et~al.}{2017}]{Gupta2017}
Gupta Y.,  et~al., 2017, CURRENT SCIENCE, 113, 707

\bibitem[\protect\citeauthoryear{Harris}{Harris}{1978}]{Harris}
Harris F.,  1978, \mn@doi [Proceedings of the IEEE] {10.1109/PROC.1978.10837}, 66, 51

\bibitem[\protect\citeauthoryear{{H{\"o}gbom}}{{H{\"o}gbom}}{1974}]{Hogbom1974}
{H{\"o}gbom} J.~A.,  1974, \aaps, \href {https://ui.adsabs.harvard.edu/abs/1974A&AS...15..417H} {15, 417}

\bibitem[\protect\citeauthoryear{{Iacobelli} et~al.,}{{Iacobelli} et~al.}{2013}]{Iacobelli2013b}
{Iacobelli} M.,  et~al., 2013, \mn@doi [\aap] {10.1051/0004-6361/201322013}, \href {http://adsabs.harvard.edu/abs/2013A%26A...558A..72I} {558, A72}

\bibitem[\protect\citeauthoryear{{Kennedy}, {Bull}, {Wilensky}, {Burba}  \& {Choudhuri}}{{Kennedy} et~al.}{2023}]{Kennedy2023}
{Kennedy} F.,  {Bull} P.,  {Wilensky} M.~J.,  {Burba} J.,   {Choudhuri} S.,  2023, \mn@doi [\apjs] {10.3847/1538-4365/acc324}, \href {https://ui.adsabs.harvard.edu/abs/2023ApJS..266...23K} {266, 23}

\bibitem[\protect\citeauthoryear{{Kern} \& {Liu}}{{Kern} \& {Liu}}{2021}]{Kern2021}
{Kern} N.~S.,  {Liu} A.,  2021, \mn@doi [\mnras] {10.1093/mnras/staa3736}, \href {https://ui.adsabs.harvard.edu/abs/2021MNRAS.501.1463K} {501, 1463}

\bibitem[\protect\citeauthoryear{Kern et~al.,}{Kern et~al.}{2020}]{Kern2020}
Kern N.~S.,  et~al., 2020, \mn@doi [The Astrophysical Journal] {10.3847/1538-4357/ab67bc}, 890, 122

\bibitem[\protect\citeauthoryear{Kolopanis et~al.,}{Kolopanis et~al.}{2019}]{Kolopanis2019}
Kolopanis M.,  et~al., 2019, \mn@doi [ApJ] {10.3847/1538-4357/ab3e3a}, 883, 133

\bibitem[\protect\citeauthoryear{Kolopanis, Pober, Jacobs  \& McGraw}{Kolopanis et~al.}{2023}]{Kolopanis2023}
Kolopanis M.,  Pober J.~C.,  Jacobs D.~C.,   McGraw S.,  2023, \mn@doi [\mnras] {10.1093/mnras/stad845}, 521, 5120

\bibitem[\protect\citeauthoryear{{Koopmans} et~al.,}{{Koopmans} et~al.}{2015}]{Koopmans2015}
{Koopmans} L.,  et~al., 2015, Advancing Astrophysics with the Square Kilometre Array (AASKA14), \href {http://adsabs.harvard.edu/abs/2015aska.confE...1K} {p.~1}

\bibitem[\protect\citeauthoryear{{La Porta}, {Burigana}, {Reich}  \& {Reich}}{{La Porta} et~al.}{2008}]{LaPorta2008}
{La Porta} L.,  {Burigana} C.,  {Reich} W.,   {Reich} P.,  2008, \mn@doi [\aap] {10.1051/0004-6361:20078435}, \href {http://adsabs.harvard.edu/abs/2008A%26A...479..641L} {479, 641}

\bibitem[\protect\citeauthoryear{Li et~al.,}{Li et~al.}{2019}]{Li2019}
Li W.,  et~al., 2019, \mn@doi [ApJ] {10.3847/1538-4357/ab55e4}, 887, 141

\bibitem[\protect\citeauthoryear{{Lonsdale} et~al.,}{{Lonsdale} et~al.}{2009}]{Lonsdale2009}
{Lonsdale} C.~J.,  et~al., 2009, \mn@doi [IEEE Proceedings] {10.1109/JPROC.2009.2017564}, \href {http://adsabs.harvard.edu/abs/2009IEEEP..97.1497L} {97, 1497}

\bibitem[\protect\citeauthoryear{{McMullin}, {Waters}, {Schiebel}, {Young}  \& {Golap}}{{McMullin} et~al.}{2007}]{casa07}
{McMullin} J.~P.,  {Waters} B.,  {Schiebel} D.,  {Young} W.,   {Golap} K.,  2007, in {Shaw} R.~A.,  {Hill} F.,   {Bell} D.~J.,  eds,  Astronomical Society of the Pacific Conference Series Vol. 376, Astronomical Data Analysis Software and Systems XVI. p.~127

\bibitem[\protect\citeauthoryear{Mellema et~al.,}{Mellema et~al.}{2013}]{Mellema2013}
Mellema G.,  et~al., 2013, \mn@doi [Experimental Astronomy] {10.1007/s10686-013-9334-5}, 36, 235

\bibitem[\protect\citeauthoryear{Mertens, Ghosh  \& Koopmans}{Mertens et~al.}{2018}]{Mertens2018}
Mertens F.~G.,  Ghosh A.,   Koopmans L. V.~E.,  2018, \mn@doi [\mnras] {10.1093/mnras/sty1207}, 478, 3640

\bibitem[\protect\citeauthoryear{Mertens et~al.,}{Mertens et~al.}{2020}]{Mertens2020}
Mertens F.~G.,  et~al., 2020, \mn@doi [MNRAS] {10.1093/mnras/staa327}, 493, 1662

\bibitem[\protect\citeauthoryear{{Mondal}, {Bharadwaj}  \& {Majumdar}}{{Mondal} et~al.}{2017}]{Mondal2017}
{Mondal} R.,  {Bharadwaj} S.,   {Majumdar} S.,  2017, \mn@doi [\mnras] {10.1093/mnras/stw2599}, \href {https://ui.adsabs.harvard.edu/abs/2017MNRAS.464.2992M} {464, 2992}

\bibitem[\protect\citeauthoryear{{Mondal}, {Bharadwaj}  \& {Datta}}{{Mondal} et~al.}{2018}]{Mondal2018}
{Mondal} R.,  {Bharadwaj} S.,   {Datta} K.~K.,  2018, \mn@doi [\mnras] {10.1093/mnras/stx2888}, \href {https://ui.adsabs.harvard.edu/abs/2018MNRAS.474.1390M} {474, 1390}

\bibitem[\protect\citeauthoryear{{Morales} \& {Hewitt}}{{Morales} \& {Hewitt}}{2004}]{Morales2004}
{Morales} M.~F.,  {Hewitt} J.,  2004, \mn@doi [\apj] {10.1086/424437}, \href {http://adsabs.harvard.edu/abs/2004ApJ...615....7M} {615, 7}

\bibitem[\protect\citeauthoryear{{Morales}, {Hazelton}, {Sullivan}  \& {Beardsley}}{{Morales} et~al.}{2012}]{Morales2012}
{Morales} M.~F.,  {Hazelton} B.,  {Sullivan} I.,   {Beardsley} A.,  2012, \mn@doi [\apj] {10.1088/0004-637X/752/2/137}, \href {http://adsabs.harvard.edu/abs/2012ApJ...752..137M} {752, 137}

\bibitem[\protect\citeauthoryear{Offringa et~al.,}{Offringa et~al.}{2015}]{Offringa2015}
Offringa A.~R.,  et~al., 2015, \mn@doi [Publications of the Astronomical Society of Australia] {10.1017/pasa.2015.7}, 32, e008

\bibitem[\protect\citeauthoryear{Olivari, Dickinson, Battye, Ma, Costa, Remazeilles  \& Harper}{Olivari et~al.}{2018}]{Olivari2018}
Olivari L.~C.,  Dickinson C.,  Battye R.~A.,  Ma Y.-Z.,  Costa A.~A.,  Remazeilles M.,   Harper S.,  2018, \mn@doi [\mnras] {10.1093/mnras/stx2621}, 473, 4242

\bibitem[\protect\citeauthoryear{Overzier, R{\"o}ttgering, Rengelink  \& Wilman}{Overzier et~al.}{2003}]{Overzier2003}
Overzier R.,  R{\"o}ttgering H.,  Rengelink R.,   Wilman R.,  2003, Astronomy \& Astrophysics, 405, 53

\bibitem[\protect\citeauthoryear{{Paciga} et~al.,}{{Paciga} et~al.}{2013}]{Paciga2013}
{Paciga} G.,  et~al., 2013, \mn@doi [\mnras] {10.1093/mnras/stt753}, \href {http://adsabs.harvard.edu/abs/2013MNRAS.433..639P} {433, 639}

\bibitem[\protect\citeauthoryear{{Pal}, {Bharadwaj}, {Ghosh}  \& {Choudhuri}}{{Pal} et~al.}{2021}]{Pal2020}
{Pal} S.,  {Bharadwaj} S.,  {Ghosh} A.,   {Choudhuri} S.,  2021, \mn@doi [\mnras] {10.1093/mnras/staa3831}, \href {https://ui.adsabs.harvard.edu/abs/2021MNRAS.501.3378P} {501, 3378}

\bibitem[\protect\citeauthoryear{Pal et~al.,}{Pal et~al.}{2022}]{Pal2022}
Pal S.,  et~al., 2022, \mn@doi [\mnras] {10.1093/mnras/stac2419}, 516, 2851

\bibitem[\protect\citeauthoryear{{Parsons} \& {Backer}}{{Parsons} \& {Backer}}{2009}]{Parsons2009}
{Parsons} A.~R.,  {Backer} D.~C.,  2009, \mn@doi [\aj] {10.1088/0004-6256/138/1/219}, \href {https://ui.adsabs.harvard.edu/abs/2009AJ....138..219P} {138, 219}

\bibitem[\protect\citeauthoryear{{Parsons}, {Pober}, {Aguirre}, {Carilli}, {Jacobs}  \& {Moore}}{{Parsons} et~al.}{2012}]{Parsons2012b}
{Parsons} A.~R.,  {Pober} J.~C.,  {Aguirre} J.~E.,  {Carilli} C.~L.,  {Jacobs} D.~C.,   {Moore} D.~F.,  2012, \mn@doi [\apj] {10.1088/0004-637X/756/2/165}, \href {http://adsabs.harvard.edu/abs/2012ApJ...756..165P} {756, 165}

\bibitem[\protect\citeauthoryear{{Parsons} et~al.,}{{Parsons} et~al.}{2014}]{Parsons2014}
{Parsons} A.~R.,  et~al., 2014, \mn@doi [\apj] {10.1088/0004-637X/788/2/106}, \href {http://adsabs.harvard.edu/abs/2014ApJ...788..106P} {788, 106}

\bibitem[\protect\citeauthoryear{Patil et~al.,}{Patil et~al.}{2017}]{Patil2017}
Patil A.~H.,  et~al., 2017, ApJ, 838, 65

\bibitem[\protect\citeauthoryear{Patwa, Sethi  \& Dwarakanath}{Patwa et~al.}{2021}]{Patwa2021}
Patwa A.~K.,  Sethi S.,   Dwarakanath K.~S.,  2021, \mn@doi [\mnras] {10.1093/mnras/stab989}, 504, 2062

\bibitem[\protect\citeauthoryear{Paul et~al.,}{Paul et~al.}{2016}]{Paul2016}
Paul S.,  et~al., 2016, \mn@doi [\apj] {10.3847/1538-4357/833/2/213}, 833, 213

\bibitem[\protect\citeauthoryear{{Planck Collaboration} et~al.,}{{Planck Collaboration} et~al.}{2020}]{Planck2020f}
{Planck Collaboration} et~al., 2020, \mn@doi [\aap] {10.1051/0004-6361/201833910}, \href {https://ui.adsabs.harvard.edu/abs/2020A&A...641A...6P} {641, A6}

\bibitem[\protect\citeauthoryear{Pober et~al.,}{Pober et~al.}{2016}]{Pober2016}
Pober J.~C.,  et~al., 2016, \mn@doi [\apj] {10.3847/0004-637x/819/1/8}, 819, 8

\bibitem[\protect\citeauthoryear{{Prabu} et~al.,}{{Prabu} et~al.}{2015}]{Prabu2015}
{Prabu} T.,  et~al., 2015, \mn@doi [Experimental Astronomy] {10.1007/s10686-015-9444-3}, \href {https://ui.adsabs.harvard.edu/abs/2015ExA....39...73P} {39, 73}

\bibitem[\protect\citeauthoryear{{Roberts}, {Lehar}  \& {Dreher}}{{Roberts} et~al.}{1987}]{Roberts1987}
{Roberts} D.~H.,  {Lehar} J.,   {Dreher} J.~W.,  1987, \mn@doi [\aj] {10.1086/114383}, \href {https://ui.adsabs.harvard.edu/abs/1987AJ.....93..968R} {93, 968}

\bibitem[\protect\citeauthoryear{{Rogers} \& {Bowman}}{{Rogers} \& {Bowman}}{2008}]{Rogers2008}
{Rogers} A.~E.~E.,  {Bowman} J.~D.,  2008, \mn@doi [\aj] {10.1088/0004-6256/136/2/641}, \href {http://adsabs.harvard.edu/abs/2008AJ....136..641R} {136, 641}

\bibitem[\protect\citeauthoryear{{Swarup}, {Ananthakrishnan}, {Kapahi}, {Rao}, {Subrahmanya}  \& {Kulkarni}}{{Swarup} et~al.}{1991}]{Swarup1991}
{Swarup} G.,  {Ananthakrishnan} S.,  {Kapahi} V.~K.,  {Rao} A.~P.,  {Subrahmanya} C.~R.,   {Kulkarni} V.~K.,  1991, Current Science, Vol.~60, NO.2/JAN25, P.~95, 1991, \href {http://adsabs.harvard.edu/abs/1991CuSc...60...95S} {60, 95}

\bibitem[\protect\citeauthoryear{{Tingay} et~al.,}{{Tingay} et~al.}{2013}]{Tingay2013}
{Tingay} S.~J.,  et~al., 2013, \mn@doi [\pasa] {10.1017/pasa.2012.007}, \href {http://adsabs.harvard.edu/abs/2013PASA...30....7T} {30, e007}

\bibitem[\protect\citeauthoryear{Trott, Wayth  \& Tingay}{Trott et~al.}{2012}]{Trott2012}
Trott C.~M.,  Wayth R.~B.,   Tingay S.~J.,  2012, \apj, 757, 101

\bibitem[\protect\citeauthoryear{Trott et~al.,}{Trott et~al.}{2016}]{Trott2016a}
Trott C.~M.,  et~al., 2016, \mn@doi [The Astrophysical Journal] {10.3847/0004-637X/818/2/139}, 818, 139

\bibitem[\protect\citeauthoryear{Trott et~al.,}{Trott et~al.}{2020}]{Trott2020}
Trott C.~M.,  et~al., 2020, \mn@doi [\mnras] {10.1093/mnras/staa414}, 493, 4711

\bibitem[\protect\citeauthoryear{{Vedantham}, {Udaya Shankar}  \& {Subrahmanyan}}{{Vedantham} et~al.}{2012}]{Vedantham2012}
{Vedantham} H.,  {Udaya Shankar} N.,   {Subrahmanyan} R.,  2012, \mn@doi [\apj] {10.1088/0004-637X/745/2/176}, \href {http://adsabs.harvard.edu/abs/2012ApJ...745..176V} {745, 176}

\bibitem[\protect\citeauthoryear{Wayth et~al.,}{Wayth et~al.}{2018}]{Wayth2018}
Wayth R.~B.,  et~al., 2018, \mn@doi [Publications of the Astronomical Society of Australia] {10.1017/pasa.2018.37}, 35, e033

\bibitem[\protect\citeauthoryear{{Wilensky}, {Morales}, {Hazelton}, {Barry}, {Byrne}  \& {Roy}}{{Wilensky} et~al.}{2019}]{Wilensky2019_SSINS}
{Wilensky} M.~J.,  {Morales} M.~F.,  {Hazelton} B.~J.,  {Barry} N.,  {Byrne} R.,   {Roy} S.,  2019, \mn@doi [\pasp] {10.1088/1538-3873/ab3cad}, \href {https://ui.adsabs.harvard.edu/abs/2019PASP..131k4507W} {131, 114507}

\bibitem[\protect\citeauthoryear{{Williams} et~al.,}{{Williams} et~al.}{2016}]{williams2016}
{Williams} W.~L.,  et~al., 2016, \mn@doi [\mnras] {10.1093/mnras/stw1056}, \href {https://ui.adsabs.harvard.edu/abs/2016MNRAS.460.2385W} {460, 2385}

\bibitem[\protect\citeauthoryear{{Yatawatta}}{{Yatawatta}}{2024}]{Yatawatta2024}
{Yatawatta} S.,  2024, \mn@doi [\aap] {10.1051/0004-6361/202449158}, \href {https://ui.adsabs.harvard.edu/abs/2024A&A...692A..31Y} {692, A31}

\bibitem[\protect\citeauthoryear{{van Haarlem} et~al.,}{{van Haarlem} et~al.}{2013}]{vanHarlem2013}
{van Haarlem} M.~P.,  et~al., 2013, \mn@doi [\aap] {10.1051/0004-6361/201220873}, \href {http://adsabs.harvard.edu/abs/2013A%26A...556A...2V} {556, A2}

\makeatother
\end{thebibliography}

\appendix
\section{Realistic Simulations}
\label{sec:sim}
The simulations used in Sections \ref{sec:mwaflag} and \ref{sec:SCF}  consider only a single grid point $g$ for which we directly simulated $\vcg(\nu)$. These simulations incorporate exactly the same frequency coverage and flagging pattern as the actual MWA data. Although these simulations are adequate to analyze the effect of  the flagged frequency channels on PS estimation, and for a preliminary quantification of the signal loss introduced by SCF, they are rather limited in that they are restricted to a single grid point and they do not include several instrumental effects like the MWA baseline distribution, baseline migration and the frequency dependence of the primary beam pattern. 

In this section, we present more realistic simulations that we have carried out to validate the SCF and check its robustness. 
We follow the methodology presented in  \cite{Chatterjee2022} and \citetalias{Chatterjee2024}, which we briefly outline here. The first step in our simulations is to generate $T(\n,\nu)$ the brightness temperature distribution on the sky. The methodology used to simulate $T(\n,\nu)$ differs depending on the component of the sky signal, as discussed in the following subsections. In all cases,  we pixelize the all-sky simulations of  $T(\n,\nu)$ using \texttt{HEALPix} \citep[Hierarchical Equal Area isoLatitude Pixelization of a sphere;][]{Gorski2005}, where we choose $N_{\rm side}=512$, which corresponds to $\ell_{max}= 1535$ and a pixel size of $6.87^{'}$. 

We calculate the simulated visibilities from $T(\n,\nu)$ using eq.~(4) of \cite{Chatterjee2022}, where we use eq.~(3) of \cite{Chatterjee2022} to model the frequency-dependent MWA primary beam pattern. The simulated visibility data have the exact baseline distribution and flagging pattern of the actual MWA observation considered here. The simulated visibilities are analyzed in exactly same way as the actual data. As earlier, we have applied SCF with a smoothing scale of $2\, {\rm MHz}$ and used TTGE to estimate the PS.

\subsection{EoR 21-cm signal}
\label{sec:HI}
We have used the EoR ($z=8$) 21-cm PS from \citet{Mondal2017} as the input model PS  $P^m (\kpp, \kpar)$  for these simulations, where we assume $P^m (\kpp, \kpar)=P^m(k)$.  Assuming the signal to be a Gaussian random field (GRF), we have used eq.~(2) and eq.~(19) of \citetalias{Chatterjee2024} to simulate the sky signal $T(\n,\nu)$, which was then used to simulate the visibilities. We have used $20$ independent realizations of the simulated signal to estimate the mean and $2\sigma$ errors for all the results presented here. 

\begin{figure}
    \includegraphics[width=\columnwidth]{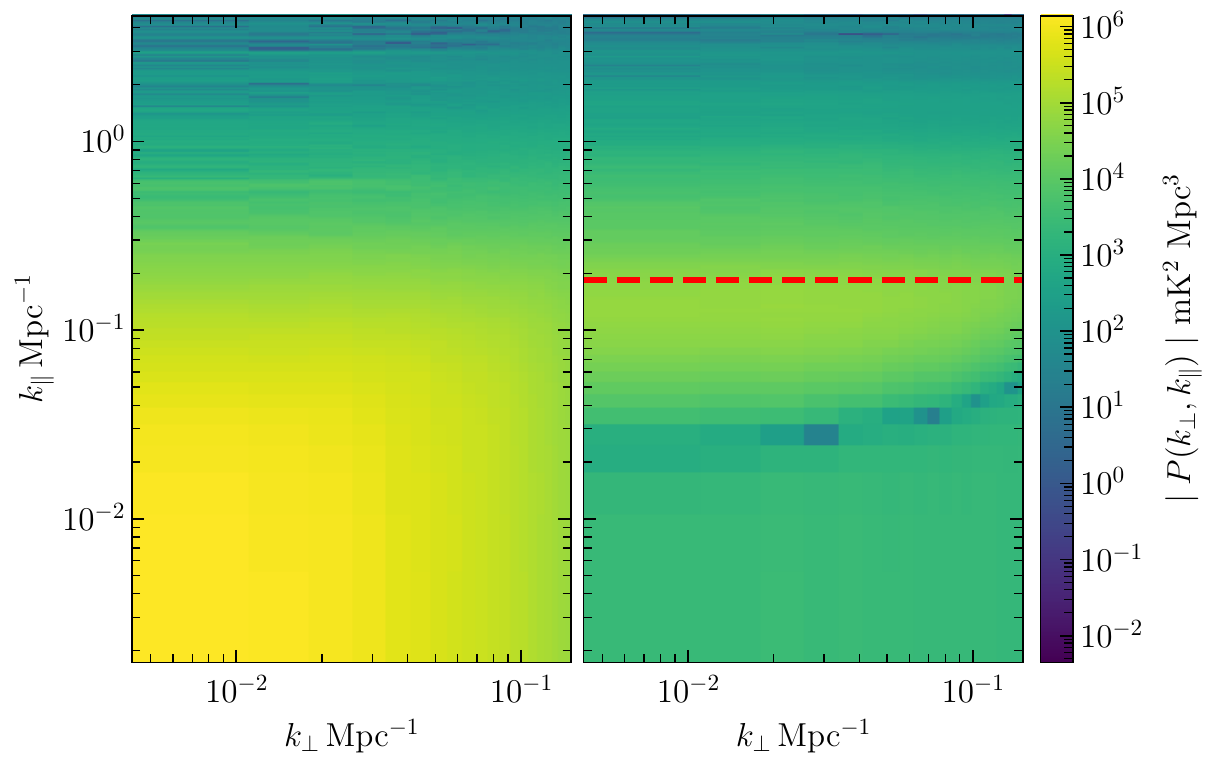}
    \caption{The cylindrical PS $\mid P(\kpp, \kpar) \mid$ of the simulated EoR 21-cm signal  before (left) and after (right) SCF. SCF filters out the smooth component of the signal leading to a significant signal loss at  $k_{\parallel}<[k_{\parallel}]_F$, which is demarcated by the red dashed line.}
    \label{fig:cylps_21}
\end{figure}
Figure~\ref{fig:cylps_21} shows the estimated cylindrical PS $\mid P(\kpp, \kpar) \mid$  before (left panel) and after (right panel) SCF. 
SCF removes the smooth component (along frequency) of the signal.
The red dashed line in the right panel shows $[k_{\parallel}]_F$ which corresponds to the 2~MHz smoothing scale used in SCF, and we expect a substantial signal loss at $k_{\parallel}\leq[k_{\parallel}]_F$. We see that,  
compared to the left panel, the right panel shows a significant drop in the values of $\mid P(\kpp, \kpar) \mid$  at $k_{\parallel}\leq[k_{\parallel}]_F$. However, we see that  $\mid P(\kpp, \kpar) \mid$  appears to remain unchanged in the range $k_{\parallel}>[k_{\parallel}]_F$ and this region can be used to estimate the EoR  21-cm PS even after SCF.  
Considering the actual data MWA data, the $(\kpp, \kpar)$ modes that we have used to estimate the 21-cm PS after SCF (Figure~\ref{fig:cylps}) are all restricted to the region $k_{\parallel} >[k_{\parallel}]_F$.

\begin{figure}
    \includegraphics[width=\columnwidth]{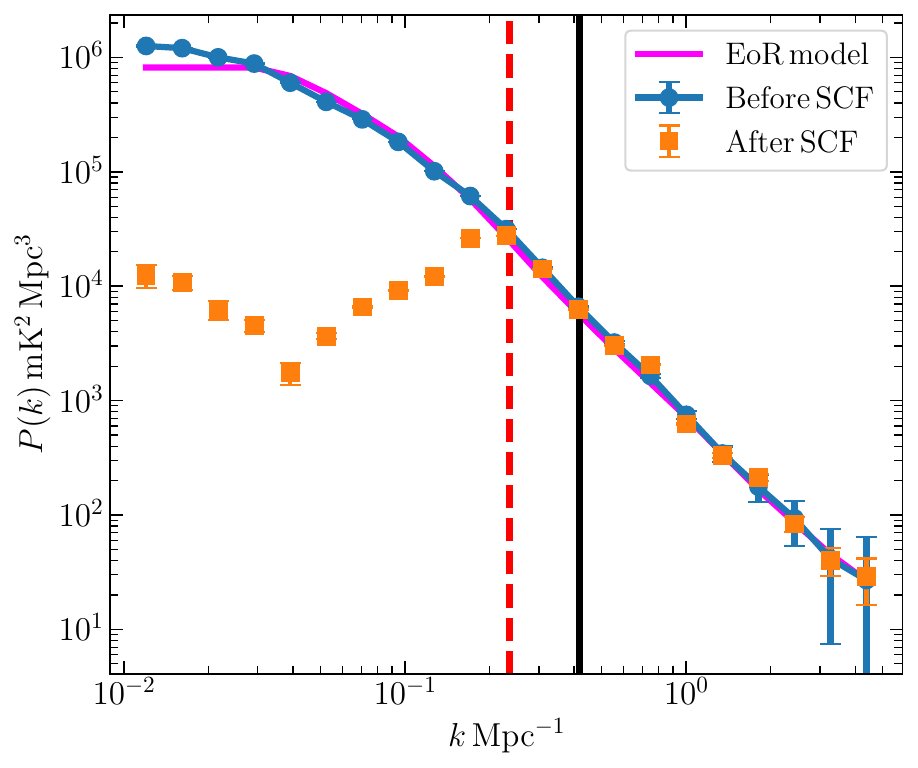}
    \caption{The estimated  $\mid P(k) \mid$ before  and after SCF. The input model EoR 21-cm PS $P^m(k)$ is also shown for reference. The red dashed vertical line shows $[k_{\parallel}]_F$. SCF introduces a significant signal loss   at $k_{\parallel}<[k_{\parallel}]_F$. Considering the actual data MWA data, we have estimated the spherical PS in the $k$-range $k \ge 0.418\,{\rm Mpc^{-1}}$, which is shown by the black vertical line.}
    \label{fig:sphps_21}
\end{figure}

The solid magenta curve in Figure~\ref{fig:sphps_21} shows $P^m(k)$  the input model EoR 21-cm PS used for the simulations, whereas the solid blue curve and the orange squares show the estimated spherical 21-cm PS before and after SCF, respectively. 
We see that SCF suppresses the power at $k<[k_{\parallel}]_F$ (red dashed line). We also see that we recover the expected 21-cm signal at $k \ge [k_{\parallel}]_F$. Our analysis in the actual MWA data are restricted to $k \ge 0.418\,{\rm Mpc^{-1}}$, shown by the black vertical line. We see that $P(k)$  estimated before and after SCF are in reasonably good agreement within the expected error bars.

\subsection{Foregrounds}
\label{sec:eps}

The most dominant contributions to the foregrounds at 154 MHz, come from the extragalactic point sources (EPS) and the diffuse Galactic synchrotron emission (DGSE). In addition to the foregrounds, the visibility data also contain system noise (NOISE) which has been kept as the same as predicted for the actual observed data (Section~\ref{sec:data}). 

The EPS are  a mix of normal galaxies, radio galaxies, quasars, star-forming galaxies, and other objects, which are unresolved by the MWA drfit-scan observation.  We use a differential source count $dN/dS$ model of the sources that is a weighted least squares $5^{\mathrm{th}}$ order polynomial fit to the 154~MHz GLEAM counts \citep{Franzen2019} and the 150~MHz counts by \cite{williams2016}, extrapolated to 154~MHz with $\alpha = -0.8$. The polynomial fit is given by
\begin{equation}
\log_{10}\left(S^{2.5} \frac{dN}{dS}\right) = \sum_{i=0}^{5} a_{i} [\log_{10}(S)]^{i} \mathrm{,}
\end{equation}
where $a_{0} = 3.52$, $a_{1} = 0.307$, $a_{2} = - 0.388$, $a_{3} = -0.0404$, $a_{4} = 0.0351$ and $a_{5} = 0.00600$ \citep{Franzen2019}. We use the fit to generate sources over the flux density range 1~mJy--50~Jy. The angular clustering of radio sources is incorporated using the angular power spectrum  $w_{\ell} \approx 1.8 \times 10^{-4} \ell^{-1.2}$ 
\citep{Overzier2003, Blake2004b, Olivari2018}. We assume that the spectral behavior of the simulated sources can be modeled as a power law $S_{\nu} \propto \nu^{\alpha}$, where for each source we randomly assign a value of $\alpha$ drawn from a Gaussian distribution with mean $-0.8$ and ${\it r.m.s.} = 0.2$ \citep{Olivari2018}.

Various observations at $150 \, {\rm MHz}$ \citep{Bernardi2009, Ghosh2012, Iacobelli2013b, Choudhuri2017} have quantified $\mathcal{C}_{\ell}$  the angular power spectrum of brightness temperature fluctuations of the DGSE. Based on these we have modelled the DGSE using 
\begin{equation}
\mathcal{C}_{\ell}(\nu_c)  = 333 \, {\rm mK}^2 \,  \left(\frac{1000}{\ell}\right)^{1.33}\,  ,
\label{eq:maps_fg}
\end{equation}
where the amplitude and the $\ell$ power-law index are from \citet{Chatterjee2025} who have analyzed the same MWA drift scan observations.
We assume that the brightness temperature  fluctuations of the DGSE are a Gaussian Random Field that follows eq.~(\ref{eq:maps_fg}) and we have used the \texttt{SYNFAST} routine of \texttt{HEALPix} 
to generate different statistically independent realizations of the brightness temperature fluctuations at $\nu_{\mathrm{c}}$. These were scaled to obtain the brightness temperature fluctuations at the other frequency channels using a spectral index $\alpha=-2.52$ in the observing bandwidth of MWA  \citep{Rogers2008}. 
Various studies indicate that the amplitude and slope have different values in different patches of the sky (\eg \citealt{LaPorta2008}, \citealt{Choudhuri2017}), and so also the spectral index \citep{DeOliveiraCosta2008}. These variations will introduce additional angular and frequency structures. However, in our simulations we have used fixed values across the entire sky. Our implementation of the EPS and DGSE closely follows \citet{Chatterjee2021}, and the reader is referred there for details.

We have simulated five realizations each for EPS and DGSE, and one realization of noise-only simulation (NOISE). The simulated visibilities have individually undergone the same analysis pipeline as the actual MWA data (DATA). In addition, we have also added the different components (EPS, DGSE and NOISE) and analyzed these. The detailed results are presented below. 

\begin{figure*}
    \includegraphics[width=0.9\textwidth]{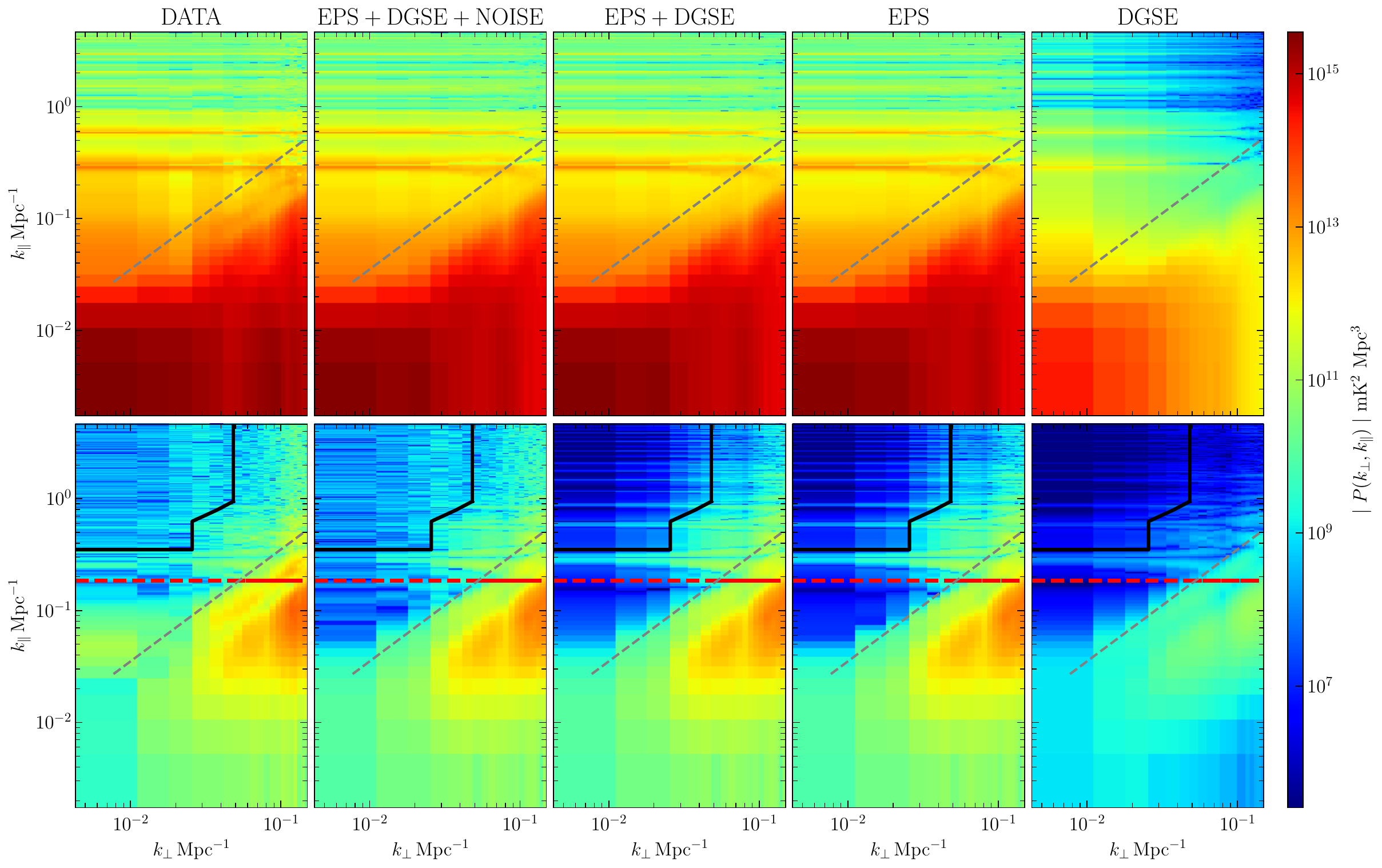}
    \caption{The top and bottom rows show the cylindrical PS $\mid P(\kpp, \kpar) \mid$ before and after SCF, respectively. Column-wise, the first column shows $\mid P(\kpp, \kpar) \mid$ from the DATA which are shown earlier in Figure~\ref{fig:cylps}. The second column shows the PS from the simulated data that has EPS, DGSE, and NOISE, and the PS fairly resembles the PS from the DATA. The third, fourth and fifth columns show the PS of the foregrounds when we do not include noise. The grey dashed curve in all the panels shows the theoretically predicted boundary of the foreground wedge. The red dashed curves in the bottom row show $[k_{\parallel}]_F$, below which  $(k_{\parallel}<[k_{\parallel}]_F)$  SCF filters out the signal.}
    \label{fig:cylps_eps}
\end{figure*}

Figure~\ref{fig:cylps_eps} shows the cylindrical PS $\mid P(\kpp, \kpar) \mid$ before (top row) and after (bottom row)  SCF, the different columns from left to right, respectively, show DATA (same as in Figure~\ref{fig:cylps}), TOTAL=EPS+DGSE+NOISE, FG=EPS+DGSE, EPS and DGSE. Comparing the first two columns, we see that the TOTAL PS from the simulation closely resembles that from the DATA, both before and after SCF. This close resemblance leads us to believe that our simulations adequately capture the dominant components of the sky signal and instrumental effects present in our observational data. We first discuss the upper row that shows the results before SCF. We see that the first four columns look very similar, by which we may say that both the simulated data and the observational data are dominated by the EPS, the DGSE being subdominant. The bulk of the foreground component is restricted within the foreground wedge, however, both the simulations and the DATA show considerable leakage in the EoR window. Note the prominent stripes that appear due to the periodic pattern of missing channels; these are present at the same values of $\kpar$ in the first four panels. The close resemblance between the second and third panels indicates that the FG PS exceeds the NOISE PS for most of the $(\kpp, \kpar)$ plane in the simulations, a conclusion that possibly also holds for the DATA PS. 

We now discuss the PS after SCF, shown in the bottom row of Figure~\ref{fig:cylps_eps}. 
For DATA, at small $\kpp$, we see that the values of $\mid P(\kpp, \kpar) \mid$  are considerably reduced after SCF as compared to before SCF. However,  we see that SCF is not equally effective at large $\kpp$  where $\mid P(\kpp, \kpar) \mid$ continues to have relatively large values even after SCF. The TOTAL foreground simulations show exactly the same behaviour, and this is also prominently visible for FG and EPS.  As discussed earlier, this arises due to baseline migration. At long baselines, or large $\kpp$, this causes the foreground signal to oscillate rapidly with frequency, and a considerable fraction of this persists after SCF. The red dashed line shows $[k_{\parallel}]_F$; we do not expect a substantial 21-cm signal loss at $k_{\parallel} > [k_{\parallel}]_F$. The black lines demarcate the region of $(\kpp, \kpar)$  space that we have used to constrain the 21-cm PS. 
Restricting our attention to this region, we see that for FG and EPS the values of $\mid P(\kpp, \kpar) \mid$  are reduced by four to five orders of magnitude  after we apply SCF. 
However, for TOTAL, the values of $\mid P(\kpp, \kpar) \mid$  are reduced by only two orders of magnitude after SCF. Our simulations thus indicate that SCF reduces the foreground PS to a level that is below the statistical fluctuations expected from the system noise present in the data. The close resemblance between DATA and TOTAL, both before and after SCF, leads us to believe that the  PS estimated from DATA is dominated by statistical fluctuations arising from noise, and this can be safely used to constrain the EoR 21-cm PS.

\begin{figure*}
    \includegraphics[width=0.9\textwidth]{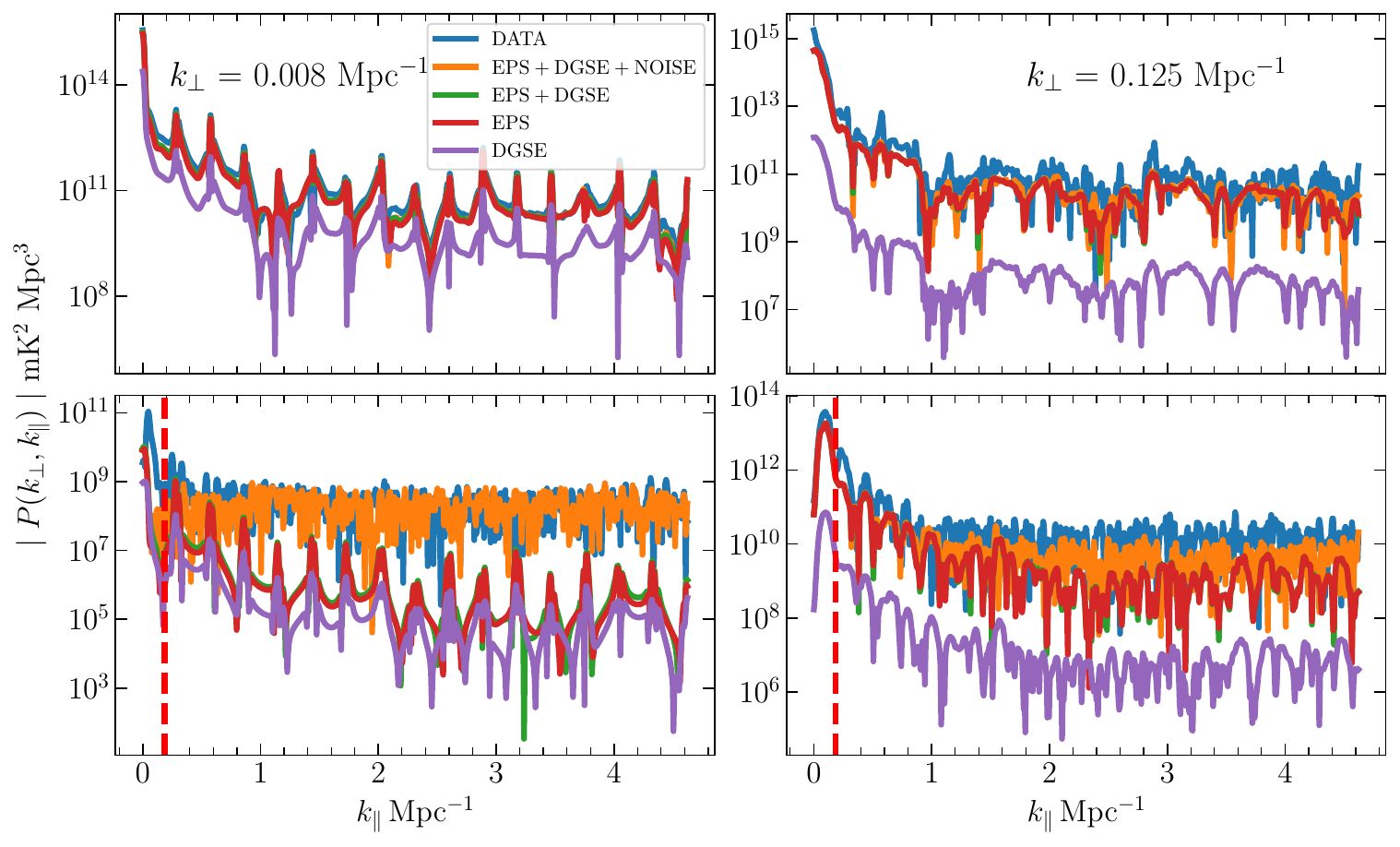}
    \caption{The power spectrum $\mid P(\kpp, \kpar) \mid$ as a function of $\kpar$ for two  different  value of $\kpp$, before (top row) and after (bottom row) SCF.  The red dashed line in the panels of the bottom row  shows $[k_{\parallel}]_F$, we do not expect significant 
    signal loss due SCF for  $k_{\parallel} > [k_{\parallel}]_F$. } 
    \label{fig:pkslice_eps}
\end{figure*}

Figure~\ref{fig:pkslice_eps} shows $\mid P(\kpp, \kpar) \mid$ as a function of $\kpar$ for two fixed $\kpp$, a small value (left) within the $(\kpp, \kpar)$ region used to constrain the EoR 21-cm PS and a large value (right) outside this region both before (top row) and after (bottom row) SCF. We see that before SCF, TOTAL closely matches DATA at small $\kpp$, whereas it is a little smaller at large $\kpp$. Note that the spikes in the simulation, which are more prominent at small $\kpp$, occur at the same $\kpar$ as in the DATA. It is also clearly visible that the TOTAL PS is entirely dominated by EPS, the NOISE and DGSE being subdominant.  After SCF, at small $\kpp$,  the FG PS drops to a level that is several orders of magnitude below TOTAL. The DATA is very close to the TOTAL PS,  which is now dominated by NOISE. We see that the foregrounds have been adequately suppressed at small $\kpp$, and we use this region to constrain the EoR 21-cm PS.  
In contrast, at large $\kpp$, the FG PS is only slightly below TOTAL. This indicates that the foregrounds have not been adequately suppressed at large $\kpp$, and we have not used these to constrain the EoR 21-cm PS.

\begin{figure}
    \includegraphics[width=\columnwidth]{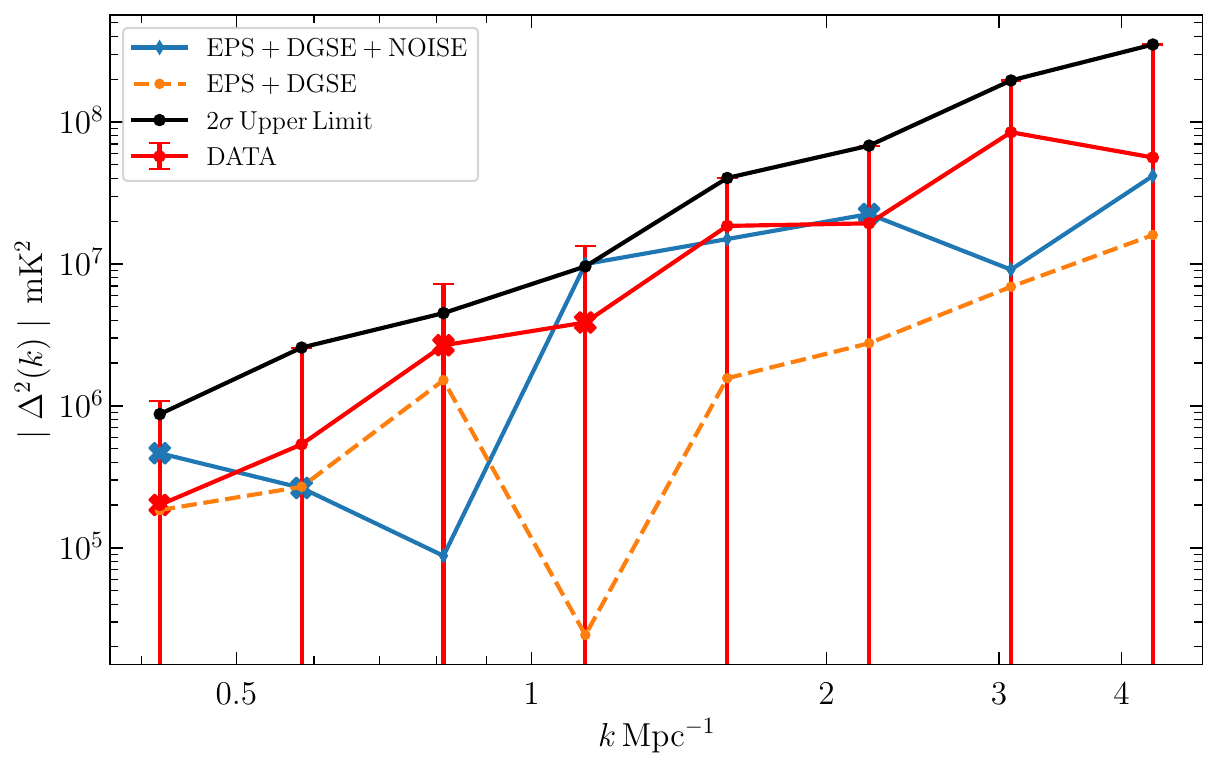}
    \caption{The  mean squared brightness temperature fluctuations $\mid \Delta^2(k) \mid$ after SCF, for both the data and the simulations. Note that the PS has both positive and negative values, and the negative values as shown with a cross for  DATA and TOTAL. We see that TOTAL PS is roughly consistent with DATA. 
    The  values $\mid \Delta^2(k) \mid$ are all well within the black solid curve that shows the $2\sigma$ upper limits.}
    \label{fig:sphericalps_eps}
\end{figure}

Figure~\ref{fig:sphericalps_eps} shows  $\mid \Delta^2(k) \mid$ the   r.m.s. brightness temperature fluctuations after SCF, for both the data and the simulations. Comparing TOTAL with FG, we see that there still is a significant contribution from NOISE even after we combine several $(\kpp, \kpar)$ modes to evaluate $P(k)$ the spherical PS. We also see that the results from TOTAL are comparable to those from DATA, and are well within the $2\sigma$ upper limits. Overall, the foreground simulations support our interpretation that SCF is able to  mitigate the artifacts due to the missing channels to a level that is below the statistical fluctuations expected from the system noise contribution.

\newpage 

\section{Choice of the window function}
\label{sec:window}

In this section, we present the rationale for choosing the Hann window $H(n)$ with a smoothing scale of $2 \, {\rm MHz}$ for SCF (Section~\ref{sec:SCF}). We first restrict ourselves to the Hann window and vary the smoothing scale to determine the optimal value.
Next, while maintaining the smoothing scale fixed at this optimal value, we study the performance of different window functions. 

\begin{figure*}
    \includegraphics[width=0.8\textwidth]{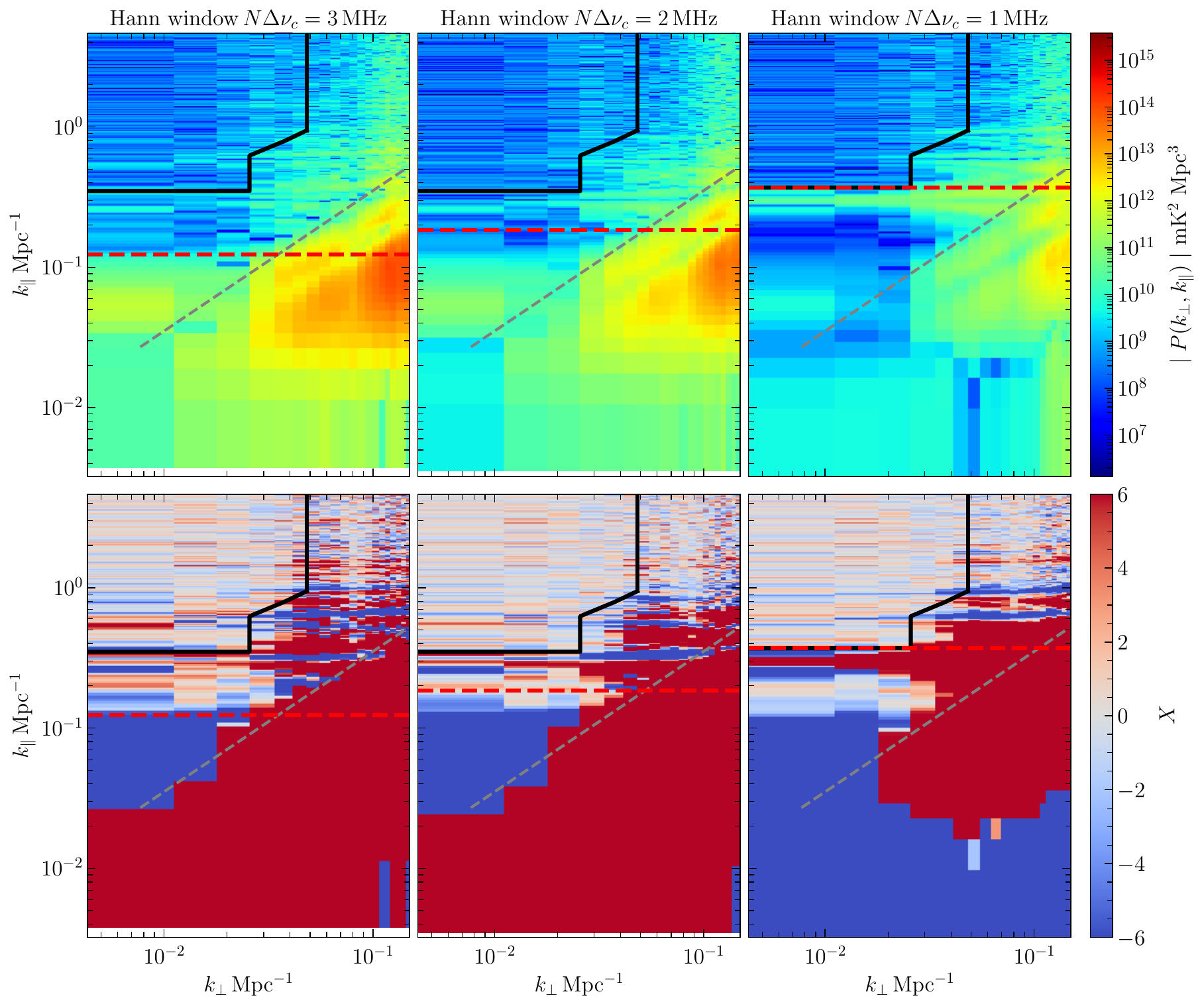}
    \caption{The cylindrical power spectrum $\mid \pk \mid $ (top row) and the heat map of $X$ (bottom row) for the Hann window function with different smoothing scales as mentioned at the top of each panel. The grey dashed curve in all the panels shows the theoretically predicted boundary of the foreground wedge. The red dashed horizontal line shows $[k_{\parallel}]_F$, below which  $(k_{\parallel}<[k_{\parallel}]_F)$  SCF filters out the 21-cm signal. The region inside the black curve is used to estimate the PS for $N \Delta \nu_c >=2 \, {\rm MHz}$. 
 The color bar for the heat map of $X$ is saturated within $\mid X \mid < 6$. }
    \label{fig:cylindricalps_N}
\end{figure*}

\begin{figure}
    \includegraphics[width=\columnwidth]{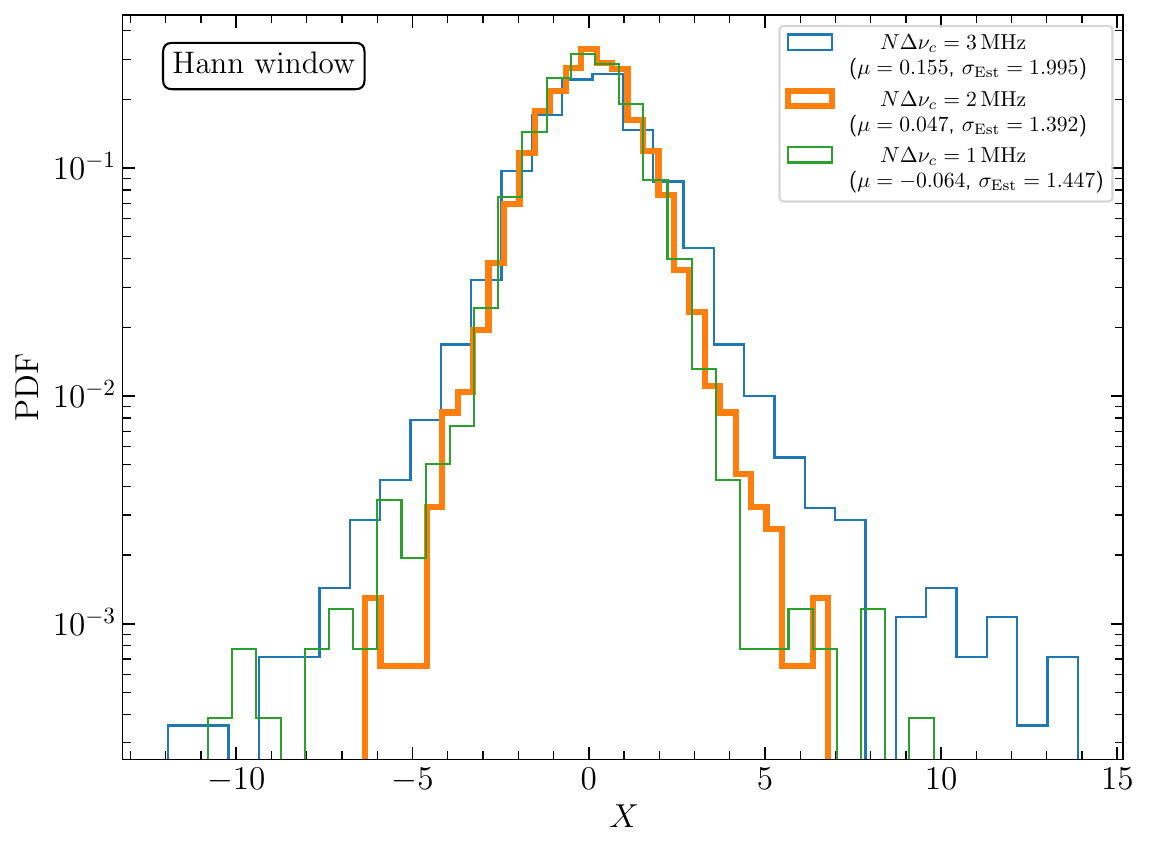}
    \caption{The PDF of $X$ (lower row of  Figure~\ref{fig:cylindricalps_N})
    for different smoothing scales.  For $N \Delta \nu_c \ge 2 \, {\rm MHz}$ , we have used the region inside the black curves in Figure~\ref{fig:cylindricalps_N}, whereas the lower boundary is decided by the red dashed horizontal line for $N \Delta \nu_c =1 \, {\rm MHz}$. }
    \label{fig:psstat_X_N}
\end{figure}

The width of the Hann window $H(n)$  is decided by $N$  (eq.~\ref{eq:Hann}), 
and here we have considered three different values $N = 25, 50\, \rm{and}\,75$, which correspond to the smoothing scales of $N\dnu_c = 1, 2\, \rm{and}\, 3~MHz$, respectively. Only the spectral features that vary faster than $N\dnu_c$ survive, and the Hann window suppresses the features that vary slower than $N\dnu_c$. We expect SCF to suppress the values of $\mid P(\kpp, \kpar) \mid$  at $k_{\parallel}<[k_{\parallel}]_F=2 \pi/(r^{'} \, N \, \dnu_c)$. Optimally, we would like to choose the largest value of $N$ for which the spikes are adequately mitigated, as the range of $\kpar$ available to estimate the EoR 21-cm signal shrinks when $N$ is reduced.  The upper row of Figure~\ref{fig:cylindricalps_N} shows $\mid P(\kpp, \kpar) \mid$ for the different values of $N\dnu_c$, and the red dashed horizontal line show the corresponding values of $[k_{\parallel}]_F$ in each panel. 
The lower row of Figure~\ref{fig:cylindricalps_N} shows the corresponding values of $X$ (eq.~\ref{eq:xstat}). For reference, in all the panels, the black curve indicates the $(\kpp, \kpar)$ region used to estimate the 21-cm signal in Section~\ref{sec:MWA}. 
Considering  $N\dnu_c =3 \,{\rm MHz}$   (left panels), we have  $[k_{\parallel}]_F = 0.092 \,{\rm Mpc}^{-1}$ for which we have a large range $k_{\parallel}<[k_{\parallel}]_F$ that can be used to estimate the PS provided the spikes have been mitigated. However, we see that the spikes are not adequately mitigated and a periodic pattern is still seen along $\kpar$ in both $P(\kpp, \kpar)$ and $X$. This is further illustrated in Figure~\ref{fig:psstat_X_N} where the PDF of $X$ shows an excess of large positive values, and also some large negative values due to the spikes. The middle panel considers $N\dnu_c =2 \,{\rm MHz}$, for which the spikes are significantly reduced. We now have $[k_{\parallel}]_F= 0.185 \, {\rm Mpc}^{-1}$, however, as discussed in Section~\ref{sec:MWA}, we have used an even smaller $\kpar$ range to estimate the 21-cm PS.  Considering  $N\dnu_c =1 \,{\rm MHz}$ (right panel), both the upper and the lower panel of  Figure~\ref{fig:cylindricalps_N}  do not indicate any significant improvement in the results. Further, the value  $[k_{\parallel}]_F=  0.370 \, {\rm Mpc}^{-1}$ exceeds the bottom boundary of the region bounded by the back curve. This reduces the $\kpar$ range available to estimate the 21-cm PS  and also increases the value of $k$ for the smallest $k$-bin, which is typically where we obtain the tightest upper limit on the 21-cm signal.   Further, considering the $X$ statistics (Figure~\ref{fig:psstat_X_N}),  the PDF of $X$ shows an excess of large positive and negative values relative to $N\dnu_c =2 \,{\rm MHz}$. The exact cause of this behaviour is not clear at present, but it could be that the narrow window function reduces the noise variance, which is reflected in the $X$ statistics.  Based on the analysis presented here, we have chosen a smoothing scale of  $2 \,{\rm MHz}$  for SCF.

\begin{figure}
    \includegraphics[width=\columnwidth]{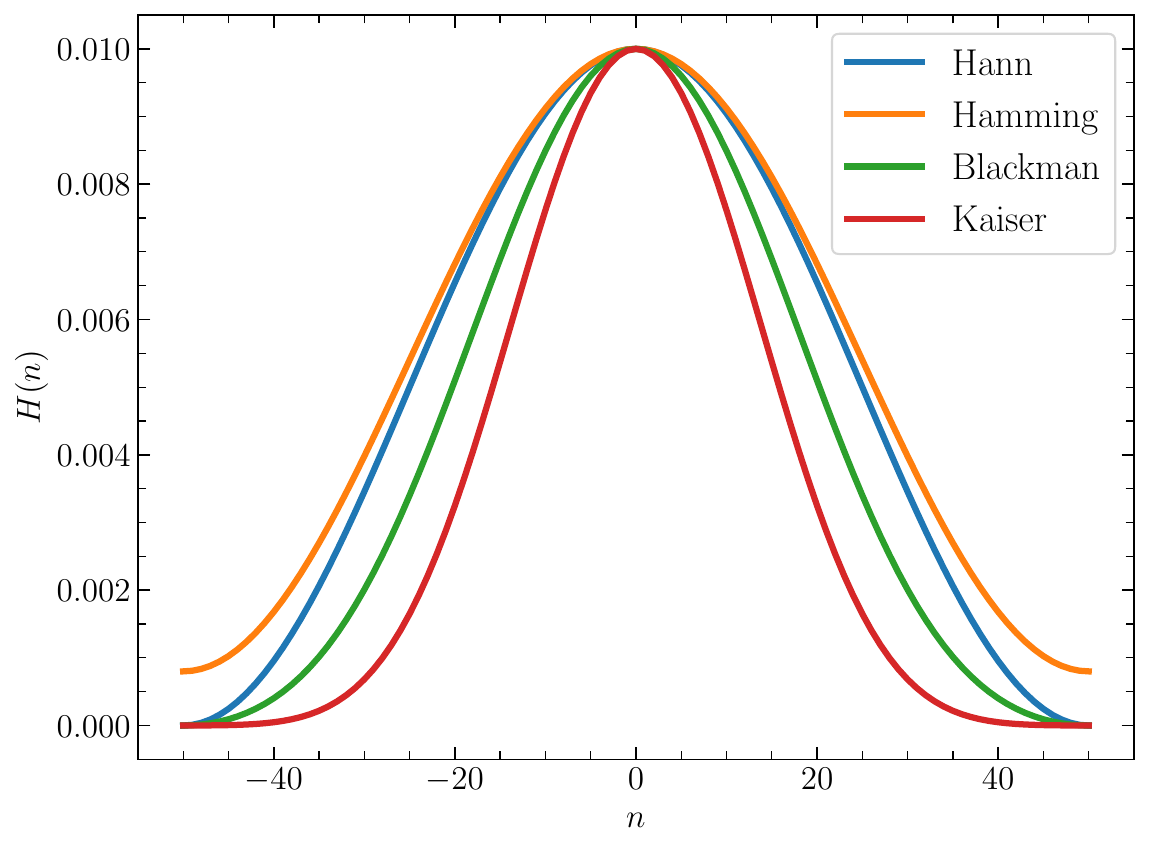}
    \caption{The window functions considered for SCF.}
    \label{fig:scf_windows}
\end{figure}

We now study the performance of different window functions,  maintaining the smoothing scale fixed at this optimal value of $2 \, {\rm MHz}$. 
We consider four different window functions,  
namely  Hann, Hamming, Blackman, and  Kaiser (see e.g. \citealt{Harris}),   which are all shown in Figure~\ref{fig:scf_windows}.   The Hann window is defined in the main text (eq.~\ref{eq:Hann}). The Hamming window is defined as
\begin{equation}
    H_{\rm{Hamming}}(n) = \frac{1}{2N} \left[ 0.54 + 0.46 \cos \left( 2 \pi \frac{n}{2N} \right) \right] , \quad -N \leq n \leq N. 
    \label{eq:hamming}
\end{equation}
Unlike the Hann  window, which goes to zero at the edges,  the Hamming window 
has a finite value of $0.08$  (Figure~\ref{fig:scf_windows}). 
The Blackman window is defined as
\begin{align}
    H_{\rm{Blackman}}(n) = & \frac{1}{2N} \left[ 0.42 + 0.5 \cos \left( 2 \pi \frac{n}{2N} \right) 
     + 0.08 \cos \left( 4 \pi \frac{n}{2N} \right) \right],  \nonumber \\
     &  -N \leq n \leq N \,.
     \label{eq:Blackman}
\end{align}
The Hamming window is broader than the Hann window, whereas the Blackman window is narrower than the Hann window.  The Kaiser window is defined as 
\begin{equation}
    H_{\rm{Kaiser}}(n) = \frac{1}{2N} \frac{I_0 \left( \beta \sqrt{1 - \left( \frac{n}{N} \right)^2} \right)}{I_0(\beta)} , \quad -N \leq n \leq N
    \label{eq:Kaiser}
\end{equation}
where $I_0(\cdot)$ is the zeroth-order modified Bessel function of the first kind, and $\beta$ controls the shape of the window. Here we have set $\beta=14$ to make  the Kaiser window narrower than the other windows considered here. Note that a value of $\beta=8.6$ results in the Blackman window.  We have performed SCF with all of these different window functions. The plots showing  $P(\kpp, \kpar)$   for the different window functions are very similar to those for the Hann window (middle panel of 
Figure~\ref{fig:cylindricalps_N}), and we have not shown these here. Note that we have used the same $(\kpp, \kpar)$  region to estimate the 21-cm PS for the different window functions.

\begin{figure}
    \includegraphics[width=\columnwidth]{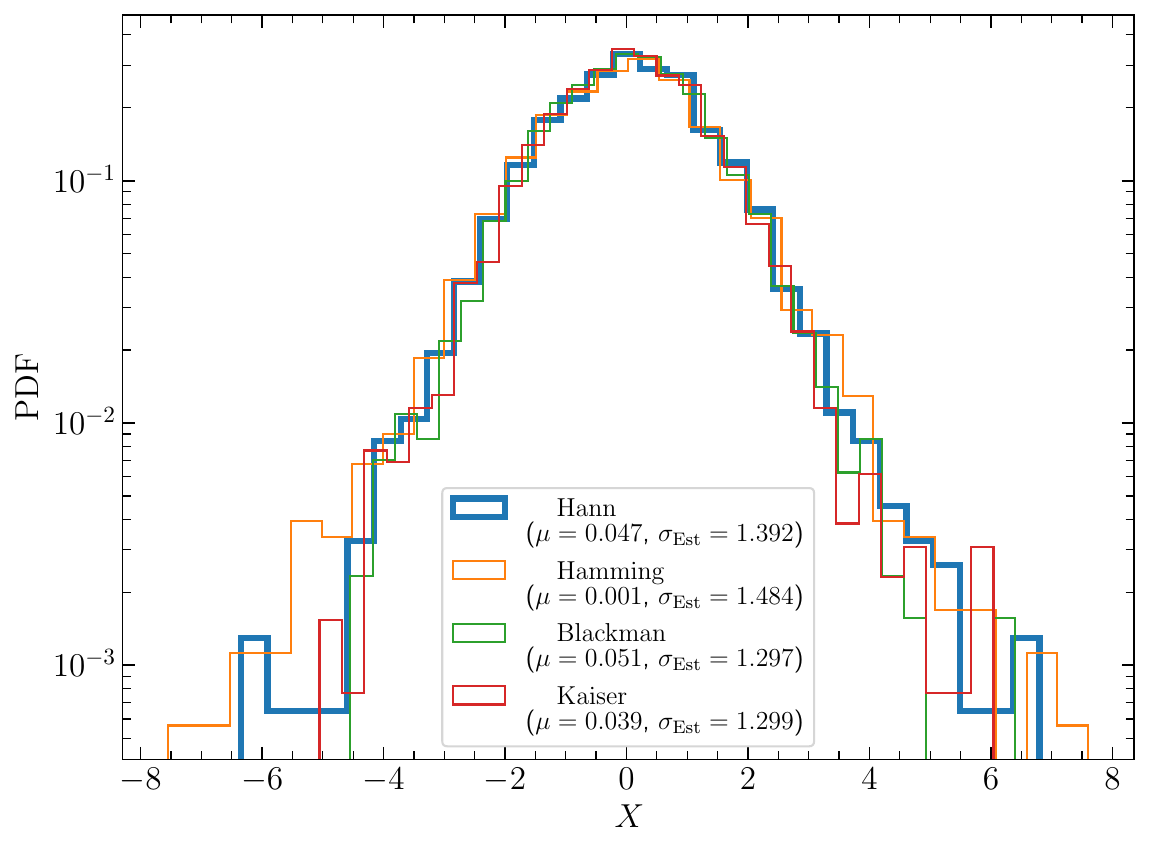}
    \caption{The PDF of $X$ using different window functions for SCF.}
    \label{fig:scf_window_X}
\end{figure}

\begin{table}
\centering
\label{tab:window}
\caption{The mean $\mu$ and the standard deviations $\sigma_{\rm Est}$ of $X$, and the $2\sigma$ $\Delta_{\rm UL}^{2}$ at $(k=0.418\,\rm{Mpc}^{-1})$ from different window functions for SCF.}
\begin{tabular}{lccc}
\hline
Window & $\mu$ & $\sigma_{\rm Est}$ & $\Delta_{\rm UL}^{2} \,\rm{mK}^2$ \\
 & & & $(k=0.418\,\rm{Mpc}^{-1})$ \\
\hline
Hann      & $0.047$ & $1.392$ & $(934.60)^2$ \\
Hamming   & $0.001$ & $1.484$ & $(965.14)^2$ \\
Blackman  & $0.051$ & $1.297$ & $(902.36)^2$ \\
Kaiser    & $0.039$ & $1.299$ & $(902.95)^2$ \\
\hline
\end{tabular}
\end{table}

Figure~\ref{fig:scf_window_X} shows the PDF of $X$ for the different window functions. We find the PDFs to be very similar, and they are all  consistent with a zero mean (Table~\ref{tab:window}).  We find the estimated excess variance $\sigma_{\rm Est}$ to be in the range $1.3 - 1.5$. Note that the Hamming window is different from the other windows, as it does not go to zero at the edges, and it produces a slightly higher value of $\sigma_{\rm Est} = 1.484$ compared to the three other window functions considered here.

\begin{figure}
    \includegraphics[width=\columnwidth]{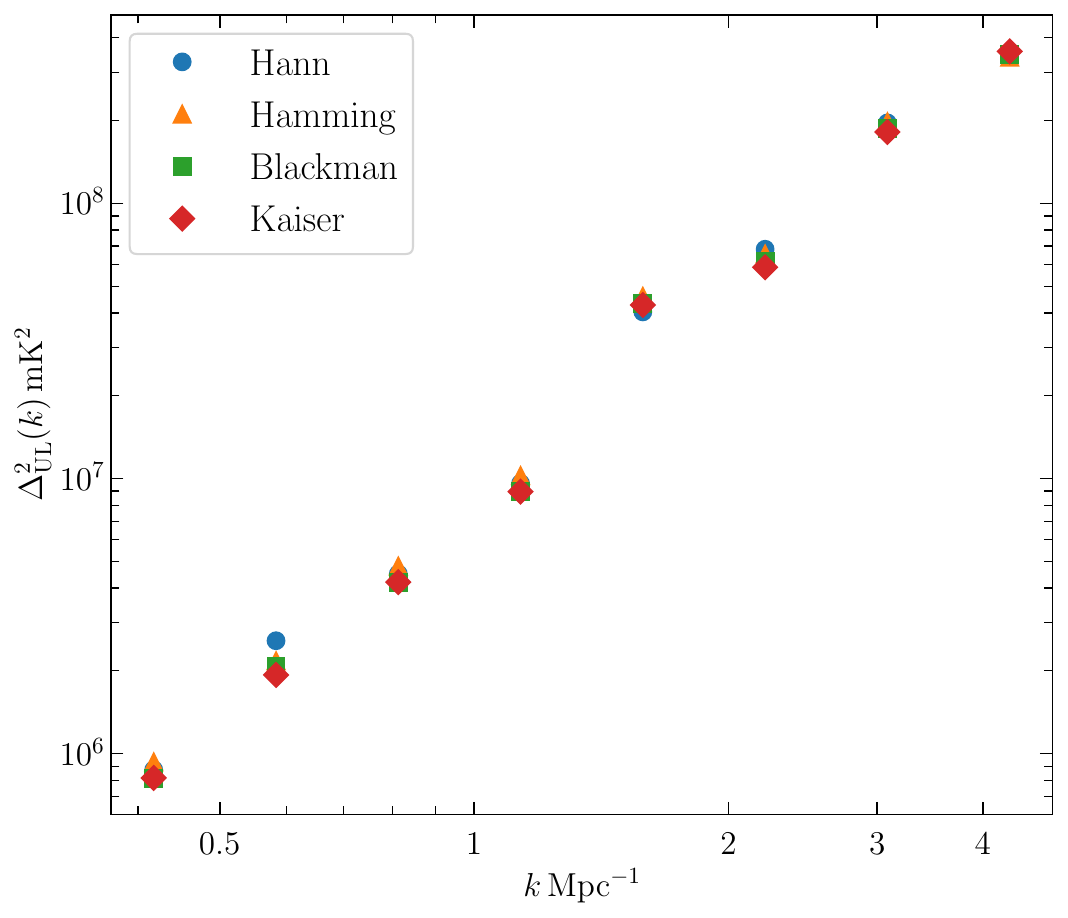}
    \caption{The $2\sigma$ upper limits for different window functions in SCF.}
    \label{fig:upperlimit_window}
\end{figure}

Figure~\ref{fig:upperlimit_window} shows the $2\sigma$ upper limits $\Delta_{\rm UL}^{2} (k)$ obtained from the different window functions. These upper limits are very similar for all the window functions considered here. The best upper limits are obtained at $k=0.418\,\rm{Mpc}^{-1}$, and we note that the upper limits obtained from the Blackman and the Kaiser windows are slightly  $(\sim 7\%)$ tighter than those from the Hann window. However, the Hann window function has a much simpler analytic form that is easier to implement, and we prefer to use this for the SCF analysis.

\bsp	
\label{lastpage}
\end{document}